\documentclass[11pt]{article}
\PassOptionsToPackage{hyphens}{url}
\usepackage[margin=0.9in]{geometry}
\usepackage{amsmath,amssymb,bm}
\usepackage{graphicx}
\usepackage{booktabs}
\usepackage{caption}
\usepackage{multicol}
\usepackage{xcolor}
\usepackage[numbers,sort&compress]{natbib}
\usepackage[colorlinks=true,linkcolor=blue,citecolor=blue,urlcolor=blue]{hyperref}
\usepackage{placeins}
\usepackage{microtype}
\graphicspath{{figs/}}
\newcommand{\rperp}{\bm{r}_{\!\perp}}
\newcommand{\kperp}{\bm{k}_{\!\perp}}
\newcommand{\shat}{\hat{\bm{s}}}

\newcommand{\Rmat}{\mathbf{R}}
\newcommand{\Cmat}{\mathbf{C}}
\newcommand{\Rs}{\mathbf{R}_{\!s}}
\newcommand{\Sn}{\Cov{n}}
\newcommand{\Cov}[1]{\mathbf{C}_{#1}}
\newcommand{\Mmat}{\mathbf{M}}

\newcommand{\Iimg}{I_{\mathrm{img}}}
\newcommand{\widefig}[2]{\makebox[\textwidth][c]{\includegraphics[width=#1\textwidth]{#2}}}

\title{\textbf{Imaging-system-aware color routers optimized for imaging information}}
\author{Hyoseok Park$^{1,2,\dagger}$, Sehyeon Park$^{3,\dagger}$, Myungjae Lee$^{3,*}$ and Yeonsang Park$^{1,2,*}$\\[2pt]
{\small $^{1}$Department of Physics, Chungnam National University,}\\[-2pt]
{\small 99 Daehak-ro, Yuseong-gu, Daejeon 34134, Republic of Korea}\\
{\small $^{2}$Institute of Quantum Systems, Chungnam National University, Daejeon 34134, Republic of Korea}\\
{\small $^{3}$Department of Materials Science and Engineering, Seoul National University,}\\[-2pt]
{\small 1 Gwanak-ro, Gwanak-gu, Seoul 08826, Republic of Korea}\\
{\small $^{\dagger}$These authors contributed equally to this work.}\\
{\small E-mail: Hyoseok Park (qkrgytjr12@gmail.com), Sehyeon Park (clairesp7@snu.ac.kr)}\\
{\small $^{*}$Corresponding authors: Myungjae Lee (myungjae@snu.ac.kr; Tel.\ +82-2-880-5819)}\\[-2pt]
{\small and Yeonsang Park (yeonsang.park@cnu.ac.kr; Tel.\ +82-42-821-6543)}\\
{\small Running title: Imaging-system-aware color routers}}
\date{}

\newcommand{\ZXYZ}{\mathbf Z}
\newcommand{\Amat}{\mathbf{A}}

\begin{document}
\maketitle

\begin{abstract}
Conventional nanophotonic color routers are typically optimized under idealized, normal plane waves. However, this standard assumption fails in real-world imaging-system environments, where structures are illuminated by converging light cones and field-dependent chief-ray angles. Here, we present an imaging-system-aware, end-to-end inverse-design framework that directly maximizes the mutual imaging information $\Iimg$ preserved by a single-layer silicon nitride color router under realistic pupil illumination. By analytically embedding the optimal reconstruction decoder directly inside the gradient loop, we co-design the optical nanostructures and the digital recovery pipeline. To scale this approach across a full sensor, we exploit the $D_4$ symmetry of the square pixel lattice, tiling $48$ distinct sensor-field regions using only six unique lithographic masks. Our optimized router is predicted to collect $2.8\times$ more photoelectrons than a conventional color-filter array. Consequently, under low-light conditions, below a green-site signal-to-noise ratio of $13.7$~dB, the color router preserves superior image information compared to the color-filter array; evaluated from its measured routing fractions together with the modeled throughput, the fabricated device reproduces this crossover at $11.1^{+2.1}_{-2.3}$~dB. This marks the first experimental demonstration, from measured routing and a modeled throughput, of a single-layer nanophotonic color router achieving a performance crossover against the color-filter array. These results establish that next-generation flat optics must shift from isolated device efficiency toward system-level co-design optimized under physical imaging-system-pupil geometry.
\end{abstract}


\vspace{1em}

\section{Introduction}
\label{sec:intro}
Nanophotonic color routers (CRs) replace the absorptive color-filter array (CFA)
\cite{bayer1976} by \emph{sorting} light by wavelength and steering each band to
its target subpixel, so that nearly all incident photons reach the photodiodes,
against $\sim\!1/3$ through dye filters. The idea has been realized as
dielectric color splitters and color-sorting metalenses
\cite{nishiwaki2013,miyata2021} and, in most of the recent
studies, free-form structures obtained by adjoint, evolutionary or
learning-based inverse design
\cite{zou2022,lee2024,rao2025,kim2024router,li2022bayer,park2026ropp}, and prototyped on commercial
submicron CMOS pixels \cite{go2025iedm}.
All of these approaches are grounded in the design framework of dielectric
metasurface-based flat optics \cite{yu2014}. The throughput
argument is compelling for photon-starved capture, where the sensor is
read-noise and shot-noise limited, not dynamic-range limited
\cite{fossum1997,fossum2014,hasinoff2010}.

\begin{figure}[tp]\centering
\includegraphics[width=\linewidth]{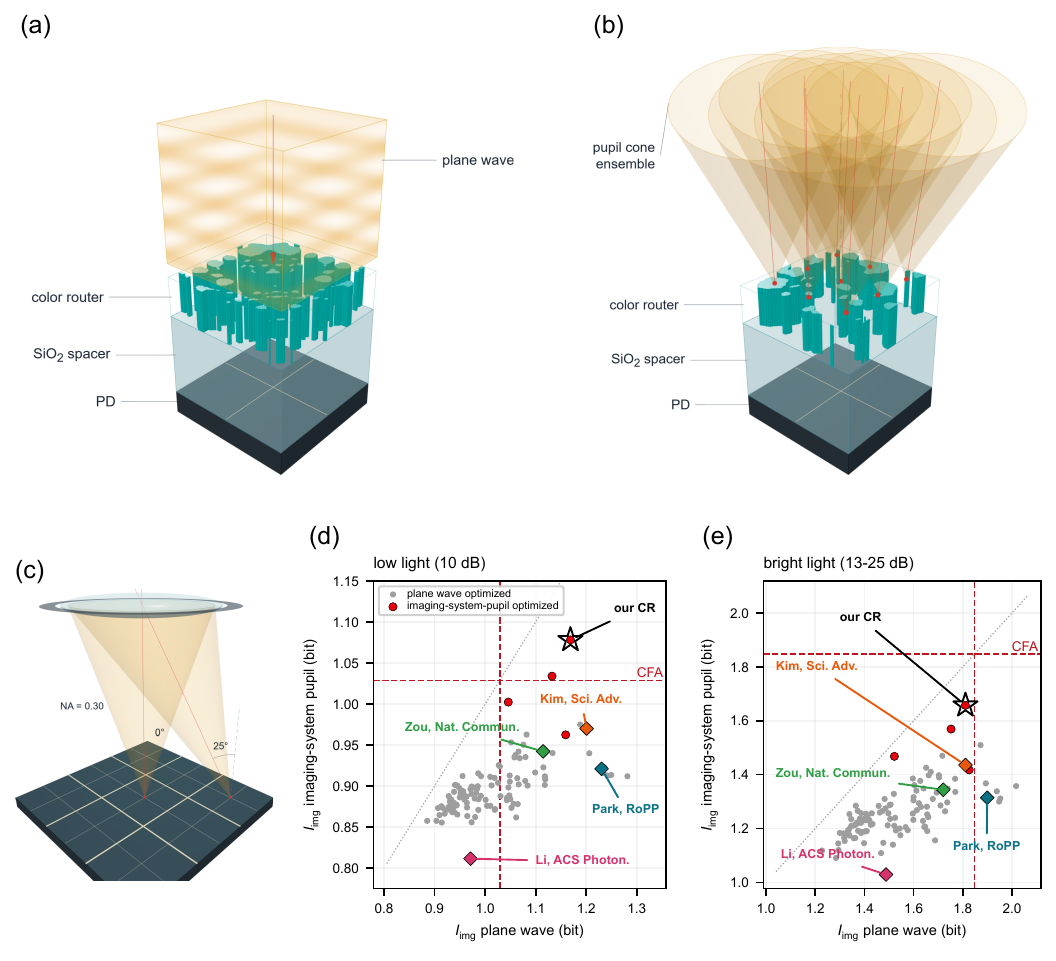}
\caption{\textbf{The design source changes the design.}
\textbf{(a,b)}~The same inverse-design loop, run under a single normal plane
wave (\textbf{a}) and under the imaging-system illumination ensemble
(\textbf{b}), produces two different free-form color routers, each a single
patterned SiN layer with four unlabeled wells in the $2\,\mu$m supercell.
\textbf{(c)}~The finite-numerical-aperture cone the ensemble is built from
($\mathrm{NA}=0.30$): normal to the sensor on axis ($0^\circ$), tilted by the
chief-ray angle at the corner ($25^\circ$ shown).
\textbf{(d)}~$109$ plane-wave-optimized single-layer color routers (grey),
imaging-system-pupil-optimized routers (red; hollow star, this work) and four
reconstructed literature routers (diamonds; orange~\cite{kim2024router},
green~\cite{zou2022}, pink~\cite{li2022bayer}, teal~\cite{park2026ropp}), scored
by their normal-plane-wave $\Iimg$ (horizontal) against their imaging-system-pupil
$\Iimg$ (vertical) in low light. Red dashed lines, the matched CFA reference on
each axis; dotted diagonal, equal information under the two sources.
\textbf{(e)}~The same designs scored over the full exposure ladder.
Literature reconstruction is validated in Sec.~S5.}
\label{fig:framework}
\end{figure}

That gain carries a cost, since two premises near-universal in the CR
literature are each violated by an imaging system.
The first premise concerns the source. A CR sorts color by angle-dependent
interference, so its spectral response is tied to the illumination geometry in a
way an absorptive filter never is. Published CR designs are optimized under
a single normal plane wave
\cite{nishiwaki2013,zou2022}. Behind a lens,
each cell instead receives a converging cone whose half-angle is set by the
numerical aperture ($\mathrm{NA}\!\approx\!0.3$ for a fast mobile lens) with
an axis normal to the sensor plane only at the optical center, tilting by the
chief-ray angle (CRA) $\theta_{\mathrm{CRA}}$ toward the corners and reaching
$\sim\!30^\circ$ for a microlens-shift-matched sensor
\cite{vaillant2004,vaillant2008,hong2025vlsi}. Representing that finite-$f$-number
illumination as an incoherent angular sum of plane waves is long established in
CMOS pixel simulation \cite{vaillant2007,vaillant2008}. Oblique incidence has
entered the CR literature only recently, and only as a tilted plane wave:
robustness to a single tilted plane wave or to an averaged range of plane-wave
angles \cite{zhao2021,jeon2025}, and per-position chief-ray tilt across the full
sensor plane \cite{kim2026stitch}, with Lee
\emph{et al.} using customized incident plane waves under non-periodic boundaries
to suppress interpixel crosstalk \cite{lee2024}. A
real pixel is lit by a converging finite-NA cone rather than a plane wave,
and an extended scene produces an incoherent
superposition of such cones integrated over the pixel. Prior work optimized
robustness over selected plane-wave angles; the design here instead enters the
physically weighted two-dimensional pupil distribution, including its
field-dependent chief-ray shift, directly into the optimization loop.
It is therefore unclear whether the plane-wave approximation merely biases the
reported performance of a color router, or changes which structure
should be designed in the first place. Here we show that it does the latter.
The design that wins under a plane wave loses under the imaging-system pupil.
We therefore reformulate the design problem, replacing plane-wave RGB-routing
efficiency with the imaging information a device's measurements preserve under
the illumination ensemble the imaging system delivers.

The second premise is that a CR must mimic a conventional CFA. Consequently,
most designs focus on maximizing physical routing efficiency into a Bayer-like
pixel grid \cite{zou2022,seo2026,miyata2026}. However, this rigid target ignores
the role of the decoder. The decoder is the computational reconstruction step
that demixes raw, overlapping measurements into standard color coordinates. Even when a
decoder is used in CR research, its role remains passive. It is typically used
as a post-hoc tool to rank finished designs \cite{johlin2021}, rather than an
active component that shapes the nanostructure during optimization.

This decoupled approach contrasts with computational imaging. In that field,
co-designing optical encoders and reconstruction pipelines is a well-established
practice \cite{lin2021e2e,sun2025}. Indeed, recovery-oriented color design has a
long history in color engineering. Examples range from filter-set goodness
measures \cite{vora1993} to joint spatio-spectral mosaic designs
\cite{hirakawa2008}. Other frameworks score optical encoders by the information
content of raw measurements rather than resemblance targets
\cite{pinkard2025,kabuli2026}. However, nanophotonic color routers have remained
isolated from these co-design principles. To our knowledge, no four-site color
router has been optimized this way. We present the first color-router framework that embeds
both a full-wave electromagnetic model and a digital decoder directly inside the
objective function. Crucially, this optimization occurs under realistic imaging-system
pupil illumination.

Typically, a color router directs predefined RGB bands to designated subpixels.
Instead, we treat it as a four-site spectral encoder. Its four raw responses
($q_1$--$q_4$) carry no intrinsic color identity. Then, we identify which four
measurements preserve the most information for image reconstruction under
realistic imaging-system illumination. Two key concepts implement this
approach. First, we integrate realistic imaging-system illumination directly
into the design loop (Fig.~\ref{fig:framework}b). For each sensor position
$(x,y)$, we model the full two-dimensional, spatially incoherent finite-NA
distribution $S_{x,y}(k_x,k_y)$ on the tilted exit pupil
(Fig.~\ref{fig:framework}c). The plane-wave responses are summed incoherently
using these physical pupil weights before evaluating any figure of merit.
Second, the digital decoder is embedded inside the optimization gradient. For
every candidate structure, we analytically solve for the optimal linear decoder
in closed form. We then backpropagate the reconstruction gradients through both
the pupil-average model and the electromagnetic solver. This allows Maxwell's
equations and the color prior to jointly guide the optimization. Together, these
concepts shift the optimization goal from mimicking band-pass filters to
maximizing information transmission. This shift fundamentally alters the
resulting physical structure: the design optimized under a simple plane wave
differs significantly from the one selected under realistic imaging-system illumination
(Fig.~\ref{fig:framework}).

This same position-dependent formulation enables us to scale these designs
across an entire sensor. Because the exit pupil shift varies in both magnitude
and azimuth, the optical response changes across the two-dimensional sensor
field. We tile this field into $48$ regions based on CRA and azimuth sectors.
Fortunately, the $D_4$ symmetry of the square pixel lattice reduces these $48$
regions to just six physical mask designs. We then execute our
imaging-system-aware optimization for each partitioned configuration to obtain
the localized nanostructures. These resulting designs, which directly embody our
system-level formulation, serve as the foundation for the performance analyses
and experimental validations presented in the following sections. Under
realistic illumination, the color router does not outperform the CFA at nominal
(bright-light) exposures; its performance is limited by spectral routing purity
rather than the total light collected. However, below a green-site
signal-to-noise ratio (SNR) of $13.7$~dB, the router preserves more color
information than the CFA. To our knowledge, this is the first single-layer
color router to achieve such a low-light performance crossover. Finally, we
fabricate this optimized router along with a plane-wave reference, confirming
that their measured routing matches our pupil-averaged forward model.

\section{Results}
\label{sec:results}

\subsection{Co-designing nanophotonic routers with imaging-system optics and information theory}
\label{sec:formulation}

At a sensor radius $r$, each pixel cell is illuminated by a converging light
cone with a numerical aperture (NA), tilted by the CRA
$\theta_{\mathrm{CRA}}(r)$. For a given pupil direction $n$ and polarization,
the simulated field deposits energy $I_{b,n}$ into site $b$. The total
transmittance, normalized to longitudinal power, is denoted as $T_n$. We
calibrate each per-direction field to the cell's total transmitted power
(Methods). Using solid-angle weights $w_n$, the vectorial Debye--Wolf field of a
single scene point is modeled as a coherent superposition of these pupil plane
waves \cite{wolf1959,richards1959,novotny2012} (Methods).

Because detection is quadratic in the electric field, the light cone focused at
a single scene point generates both diagonal and cross terms in $n$. However,
real-world scenes are spatially incoherent and locally uniform. As the focus
position is averaged over multiple cell periods, the off-diagonal phase terms
cancel out. This leaves only the incoherent aggregate intensity. By contrast,
summing the pupil components as fields would retain phase cross-terms
(Fig.~S2); since the spatial incoherence of extended scenes naturally averages
out these terms, any design optimized and certified purely against a
coherent-cone sum fails to transfer to a physical imaging system. The
surviving quantity is the incoherent aggregate
\begin{equation}
\boxed{\;\overline P_b(\lambda;x,y)=\sum_n w_{n,x,y}\,P_{b,n}(\lambda),
\qquad
P_{b,n}=T_n\,\frac{I_{b,n}}{\sum_{c=1}^{4}I_{c,n}}.\;}
\label{eq:incoh}
\end{equation}
Here $P_{b,n}$ acts as a cell-total-power-calibrated site-partition proxy, and
the total $T_n$ is power-normalized, while its allocation among the four sites
uses an electric-energy partition in place of a per-well Poynting-flux or
carrier-generation calculation. Equation~\eqref{eq:incoh} performs a quadrature
of the scene radiance over the shifted pupil. This formulation becomes exact
under three conditions: (i) the scene is incoherent and uniform over a patch
larger than the coherence radius ($0.61\lambda/\mathrm{NA}=1.12\,\mu$m at
$550$\,nm), (ii) the lens is shift-invariant over the cell, and (iii) the pixel
array consists of identical periodic cells. The full derivation of this
cross-spectral density, including the van Cittert--Zernike coherence scale
\cite{vancittert1934,zernike1938,wolf1954} and finite-support residues, is
detailed in Sec.~S1, where Fig.~S2 contrasts this incoherent intensity sum with
a coherent field sum.

The single, pre-computed weight set $\{(\theta_n,\phi_n),w_n\}$ accounts for
cone diffraction, aperture convolution, neighboring-cell spill, and chief-ray
tilt. Convergence of this $16$-ray quadrature rule is rigorously verified
against a dense $100$-ray reference in Sec.~S1 (Fig.~S1), demonstrating a
worst-case well-fraction shift of less than $5\times10^{-3}$. This weight set is
independent of the color router geometry and is computed only once. To model a
reference lens, we normalize these weights using a relative-illumination model
$\cos^4\theta_{\mathrm{CRA}}$ \cite{vaillant2008,catrysse2000qe,huo2010}. The
chief-ray tilt rigidly shifts every pupil sample on the disk (Methods):
\begin{equation}
(s_{x,n},s_{y,n})\;\mapsto\;(s_{x,n},s_{y,n})+\sin\theta_{\mathrm{CRA}}
\bigl(\cos\phi_{\mathrm{CRA}},\,\sin\phi_{\mathrm{CRA}}\bigr).
\label{eq:tiltshift}
\end{equation}
Thus, the illumination source is fully defined by two field coordinates: the
tilt magnitude $\theta_{\mathrm{CRA}}$ and the azimuth $\phi_{\mathrm{CRA}}$.

As a passive four-site spectral encoder, the color router's four responses
carry no intrinsic R, G, or B identities. The physical quadrant layout of these
four detector sites ($q_1$--$q_4$) and the predominant spectral band each site
collects are illustrated in Fig.~S7. The optimization objective does not
penalize or reward resemblance to traditional dye band-pass filters. Let
$\mathbf s$ represent a reflectance spectrum on our evaluation grid. While the
target colorimetric values $\Cmat\mathbf s$ are calculated using the CIE
color-matching functions \cite{cie0152018}, the sensor converts the physical
site responses into expected photoelectrons \cite{vereecke2015} (Methods). This
conversion accounts for the illuminant's spectral power distribution and a
typical backside-illuminated silicon quantum efficiency (Methods), yielding the
electron operator $\Rmat_\varepsilon$. The resulting four-site signal is:
\begin{equation}
\mathbf y=\Rmat_\varepsilon\mathbf s+\mathbf n,
\label{eq:signal}
\end{equation}
where $\mathbf n$ represents noise with a signal-dependent covariance $\Sn$
containing both shot noise and read noise. Let $\mathbf X$ denote the
zero-mean spectral latent of which $\mathbf s$ is a realization and $\mathbf Z$
the corresponding CIE XYZ target; their joint second moments are
$\Cov{\mathbf X}$, $\Cov{\mathbf XZ}$ and $\Cov{Z}$, and $\Mmat$, constructed
from $\Rmat_\varepsilon$, maps $\mathbf X$ to the four-site measurement
$\mathbf Y$. Given the spectral second moment
$\Rs$ of our reflectance prior \cite{babelcolor2012}, we can solve for the
optimal linear decoder in closed form. This decoder is re-solved at every
optimization step, allowing us to differentiate the geometry directly through
the decoder rather than relying on a fixed, pre-trained one.

Our design objective is the imaging information,
$\Iimg\equiv\tfrac14 I(\mathbf Y;\mathbf Z)$. This represents a quarter of the
mutual information between the four noisy raw measurements $\mathbf Y$ and the
target CIE XYZ image $\mathbf Z$. We divide by four because the four-site cell
spans four raw pixels. Under a Gaussian noise approximation, this information
has a closed-form expression (derived in Sec.~S3):
\begin{equation}
  \Iimg=-\frac{1}{8}\sum_{i=1}^{3}\log_2\rho_i,
  \label{eq:iimg}
\end{equation}
where the $\rho_i$ are the three generalized eigenvalues of the posterior color
covariance $\Cov{Z|\mathbf Y}$ against the prior color covariance $\Cov{Z}$,
solved via $\det(\Cov{Z|\mathbf Y}-\rho\,\Cov{Z})=0$. The posterior covariance
is given by:
\begin{equation}
  \Cov{Z|\mathbf Y}
  =\Cov{Z}
  -\Cov{Z\mathbf X}\Mmat^{\!\top}
  \left(\Mmat\Cov{\mathbf X}\Mmat^{\!\top}+\Sn\right)^{-1}
  \Mmat\Cov{\mathbf XZ},
  \label{eq:posterior}
\end{equation}
where $\Mmat$ is the measurement matrix mapping the scene latent state into
expected charge, and $\Sn$ is the shot-plus-read noise covariance. Physically,
each $\rho_i\in(0,1]$ represents the fraction of prior color variance that
remains unresolved along one principal direction of the color space after
measurement. Consequently, $-\tfrac12\log_2\rho_i$ is the number of bits
delivered about that direction at the cell level. The sum runs over three
dimensions because the target color space XYZ is three-dimensional. Thus, the
sum measures the log-volume reduction in color uncertainty. A device that
measured nothing would leave all $\rho_i=1$, yielding $\Iimg=0$. All $\Iimg$
values are reported in bits per raw pixel, allowing direct comparison with a
CFA on the same sensor.

This mutual information objective offers several key mathematical advantages.
First, $\Iimg$ is invariant under any invertible linear mapping of the four
measurements. This property ensures the optimizer is never biased toward
mimicking arbitrary R, G, or B bands; additionally, site permutations during
deployment carry no penalty. Second, by analytically marginalizing the optimal
linear decoder, we avoid the need to train a reconstruction network at each
iteration. Third, because $\Iimg$ is an information metric rather than an error
metric, we can aggregate performance across an exposure ladder without
arbitrarily choosing a reconstruction-quality threshold. This ladder consists
of five reference-CFA-green levels, $\{20,40,80,160,320\}$~e$^-$ against a
$1000$~e$^-$ reference, entering with equal weight. This exposure-ladder
averaging prevents the optimizer from specializing to a single noise regime. It
forces the framework to simultaneously optimize photon collection under
read-noise limits and spectral conditioning under shot-noise limits, so that the
design is balanced across the exposure range rather than tuned to a single
light level. The nominal exposure refers
to this ladder at unit scale, corresponding to a mean green-site charge of
$124$~e$^-$, to which all exposure and SNR figures are referenced. The
optimization schedules and the noise model are specified in Secs.~S3 and~S6.
The CFA baseline uses a broad primary-pigment transmission stand-in constructed
from the published dye-filter literature, together with an $85\%$ microlens
efficiency \cite{catrysse2002,agranov2003,vaillant2004,vaillant2008,huo2010},
four quarter-area sites, and the same illuminant, relative illumination and
quantum efficiency as the CR.

\subsection{Plane-wave optimization can select the wrong design}
\label{sec:mismatch}

We evaluated $109$ single-layer color routers optimized under the conventional
normal plane wave. These candidate structures spanned a wide range of layer
heights, router-to-detector gaps, and random optimization seeds. We then scored
each design under both the idealized plane wave and our realistic, on-axis
imaging-system pupil illumination. Under a photon-starved exposure of
$12$~e$^-$ per green site ($10$~dB SNR), many plane-wave-optimized designs
comfortably outperform the CFA reference under plane-wave
illumination (Fig.~\ref{fig:framework}d). However, none of these designs clear
the CFA reference when re-evaluated under realistic imaging-system-pupil illumination
(Fig.~\ref{fig:framework}d). This performance degradation persists when the
imaging information $\Iimg$ is averaged over the full exposure ladder
($13$--$25$~dB SNR) (Fig.~\ref{fig:framework}e). All $109$ candidate designs
fall below the diagonal of equal performance, proving that plane-wave
evaluations systematically overestimate the preserved color information.

Crucially, switching to imaging-system-pupil illumination does not merely shift
performance uniformly; it fundamentally alters the ranking of the designs
(Fig.~\ref{fig:framework}d,e; Fig.~S9). The resulting crossing of the
performance lines in Fig.~S9 proves that the plane-wave ranking carries only
limited predictive information about actual imaging-system performance. For example,
the champion plane-wave design (scoring $2.02$~bits) collapses to just
$1.36$~bits under the pupil ensemble, where it is easily beaten by a design
that scored only $1.83$~bits under plane-wave conditions. Because this
performance drop varies drastically from $0.19$ to $0.66$~bits depending on
the geometry, it cannot be corrected by a constant offset. This localized
collapse is also highly evident at the spectral response level for
mirror-symmetric designs (Fig.~S5). This vulnerability is not unique to our
structures; four state-of-the-art color routers reconstructed from the
literature \cite{zou2022,kim2024router,li2022bayer,park2026ropp} exhibit the
same severe plane-to-pupil degradation (Fig.~\ref{fig:framework}d,e; Fig.~S4).
Because the imaging-system-pupil score dictates real-world imaging-system
performance, a router can only replace a CFA if it exceeds the CFA reference
under realistic pupil illumination. By this strict metric, our
information-optimal design preserves more imaging information than any
single-layer router tested, including the four reconstructed literature
designs. In the photon-starved, low-light regime, our design is the only one
that successfully surpasses the CFA baseline (Fig.~\ref{fig:framework}d).

\subsection{Design and $D_4$ deployment}
\label{sec:deployment}

Every design here is one free-form binary pattern in a single SiN
layer on a $2\,\mu$m four-site cell, exposed to air above an oxide substrate,
at a common height $h=600$\,nm, requiring one deposition and one etch for the whole
wafer, with the regions differing only in the lithography pattern, which is
position-dependent in any case. A back-illuminated CMOS stack would ordinarily
planarize such a layer in oxide
\cite{wakabayashi2010,fossum2014}, dropping the index
contrast from $2.0/1.0$ to $2.0/1.46$ and weakening the achievable routing. However, the
designs reported here keep the air interface above the pattern. Integrating
them in a planarized stack would mean recovering that contrast with height or
index. Each
pupil plane wave is one rigorous coupled-wave analysis solve
\cite{moharam1981,moharam1995a,moharam1995b,li1996a,li1996b} in the differentiable GPU implementation TORCWA \cite{kim2023}. We parameterize the pattern as a $44\times44$ latent field, upscaled to the
$128\times128$ raster through a conic filter of radius $45$\,nm and a tanh
projection in the robust erosion/dilation formulation \cite{sigmund2007,wang2011,zhou2015}, so the
latent pitch and the filter scale coincide at $45$\,nm and the free parameters
number $1936$, far fewer than the $16384$ raster cells
(Fig.~S6). Off-diagonal design azimuths
break the $y=x$ mirror of the cell, so every off-diagonal response uses the full
two-polarization path. The patterns themselves are left unconstrained for the
same reason, because away from the axis the incident ensemble is already asymmetric, so
a mirror imposed on the pattern matches no symmetry of the problem and only
removes degrees of freedom. Detector-plane conventions, the fill-factor assumption,
the $(8,8)$ differentiable surrogate and the hard-mask evaluation settings are
specified in Sec.~S2.

The chief-ray tilt (Eq.~\eqref{eq:tiltshift}) displaces the whole pupil
disk by a vector with a
magnitude \emph{and} a direction, so the effective source lives on a
two-dimensional domain
$(\theta_{\mathrm{CRA}},\phi)\in[0,25^\circ]\times[0,360^\circ)$,
the domain covered by the architecture developed
below. The CFA is insensitive to that second coordinate, because dye
absorption is isotropic to first order, so a Bayer CFA is azimuthally
flat by construction. A CR responds to it, for two reasons: it sorts color
by interference between scattered orders whose relative phases are set by the
incident direction, and the square four-site lattice it sorts \emph{into} is
itself anisotropic. At $\phi=45^\circ$ the cone axis runs along one site
diagonal, at $135^\circ$ along the other, and at $0/90^\circ$ along the cell
edges, so the same fixed pattern and site assignment meet different incidence
geometries as $\phi$ turns.

Carrying one mask across the field is expensive. The on-axis design,
re-evaluated unchanged at the six design
slots, falls from $\Iimg=1.63$ on axis to $1.33$ at twenty degrees, and its
deficit against the CFA widens from $0.22$ to $0.41$\,bit. A family
designed along a single diagonal degrades the same way in azimuth. No
single cell covers $\theta_{\mathrm{CRA}}\in[0,25^\circ]$, and the field must be
partitioned in both coordinates.

The partition uses rings $0$--$5$, $5$--$15$ and $15$--$25^\circ$ in chief-ray
angle with design centers $2.5/10/20^\circ$, crossed with azimuth sectors whose
widths narrow outward, with eight $45^\circ$ sectors in ring~1, sixteen
$22.5^\circ$ sectors in ring~2 and twenty-four $15^\circ$ sectors in ring~3,
with the sector \emph{edges} placed on the $0/45/90^\circ$ symmetry axes so
that no sector straddles one. The grading is set by the geometry, because an azimuth
mismatch displaces the pupil disk along a chord that grows with chief-ray
angle, so the adopted half-widths $22.5/11.25/7.5^\circ$ hold the
center-to-edge displacement below $15\%$ of the pupil radius in every ring,
against the $44.5\%$ a uniform $45^\circ$ sector would produce at ring~3
(Sec.~S4).

\begin{figure}[tbp]\centering
\widefig{0.92}{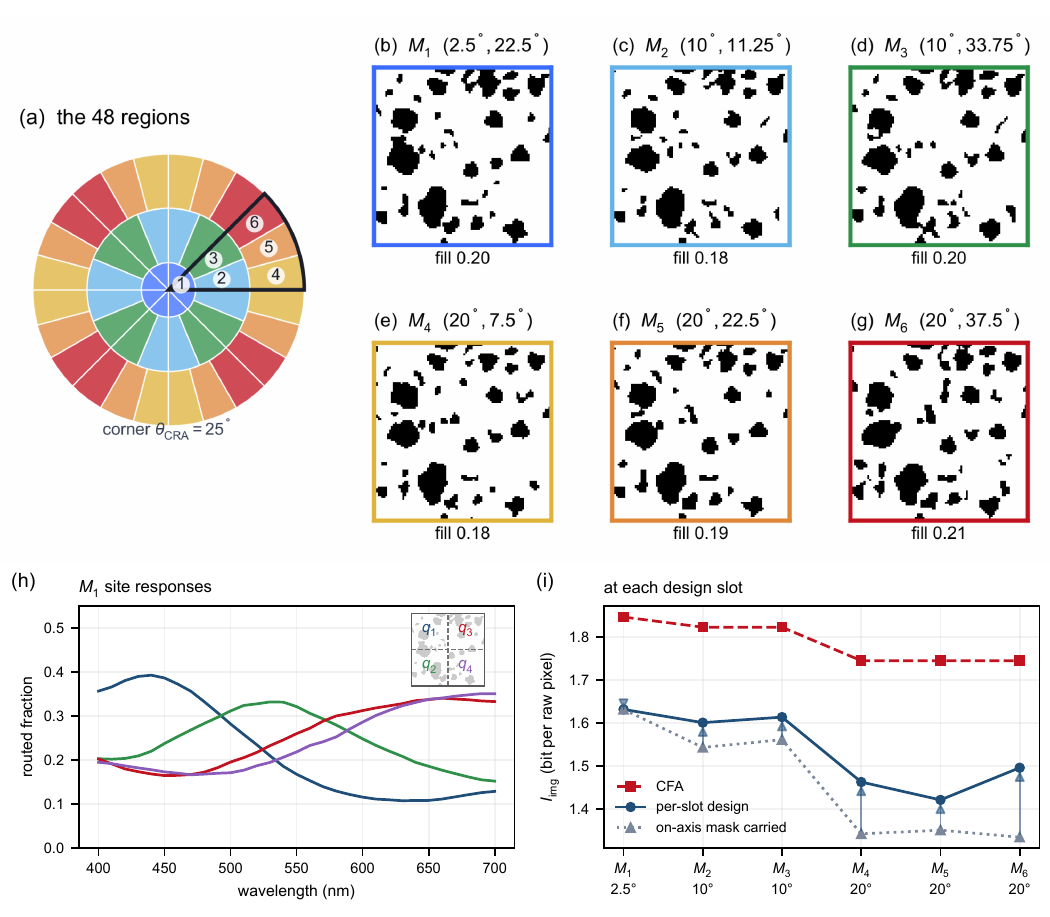}
\caption{\textbf{The six-mask designed set and what per-sector design buys.}
\textbf{(a)}~The $48$ field regions, three
rings in chief-ray angle carrying eight, sixteen and twenty-four azimuth
sectors, colored by the design that serves each. The outlined $45^\circ$
wedge is what must be designed, and the numbered markers are the six design
azimuths. \textbf{(b--g)}~The binary hard masks (SiN black, air white), labeled
with the chief-ray angle and azimuth they were designed for and with the SiN
fill fraction of the array shown.
\textbf{(h)}~Site responses of $M_1$, the inset marking the four sites on the mask;
all six designs are in Fig.~S8.
\textbf{(i)}~Imaging information at the six design slots, ordered by chief-ray
angle: the CFA simulated there, the mask designed for that slot, and
the on-axis mask carried to it unchanged, with arrows marking the gain.}
\label{fig:sector_designs}
\end{figure}

Only the $45^\circ$ first-quadrant wedge must be designed. The eight isometries
of the square lattice map the four-site cell onto itself, so applying one of
them to a pattern gives a solution of Maxwell's equations at the mapped
azimuth with the same fields relabeled. The only consequence is a static
permutation of the four sites, and that costs nothing because the
sites carry no color filters. A Bayer-labeled cell would keep only two of the
eight, the identity and the mirror through its red and blue sites, because
every other isometry moves at least one color onto a differently labeled
site; the $180^\circ$ rotation, for instance, exchanges red and blue.
Covering the field under fixed labels would therefore take four times as many
independently refined sectors. The orbit of a design azimuth is not a uniform
$45^\circ$ comb, so the wedge centers were chosen so that their orbits tile
each ring exactly. The arithmetic and the conditions under which the improper
elements apply are given in Sec.~S4. The six wedge
designs $\mathcal M_1$--$\mathcal M_6$ of Fig.~\ref{fig:sector_designs}
serve all $8+16+24=48$ regions, with no azimuth served off a design axis by
more than its sector half-width, and fully cover a sensor whose corner
chief-ray angle is $25^\circ$. A larger corner would require a fourth ring.

Exact deployment under $D_4$ still leaves each sector to be designed on its own,
because the symmetry transports a pattern to a mapped azimuth without keeping it
optimal as the chief-ray direction moves. Each of the six was therefore refined at its own slot,
starting from the carried on-axis mask, using hard-mask acceptance; the slot
azimuths are $22.5^\circ$ for $\mathcal M_1$, $11.25$ and $33.75^\circ$ for
$\mathcal M_2$ and $\mathcal M_3$, and $7.5$, $22.5$ and $37.5^\circ$ for
$\mathcal M_4$ to $\mathcal M_6$. Figure~\ref{fig:sector_designs} shows the six
masks, the $48$-region field map they tile, the site responses of
$\mathcal M_1$, and the imaging information at the
six slots. Designing each sector separately improves the weakest field positions as
well as the field-averaged performance. The six per-slot designs hold the
deficit against the CFA between $0.22$ and $0.32$\,bit at every slot (Fig.~\ref{fig:sector_designs}). As a result, the worst position on the
sensor improves by $0.087$\,bit and the spread across the field narrows from
$0.20$ to $0.11$\,bit. Spatial uniformity of color response is a separate
requirement from mean performance in an imaging system, and it is the one that microlens
shifting and chief-ray-angle matching already exist to meet
\cite{vaillant2008,catrysse2000qe}.

The mean gain is $0.09$\,bit, area-weighted over the covered field. It is zero
on axis by construction, since every run was warm-started from the carried mask
and floored at it, and it grows with chief-ray angle, from $0.05$ to $0.06$\,bit in
ring~2 and $0.07$ to $0.16$\,bit in ring~3. Within ring~3 the three design
azimuths differ by more than a factor of two: the azimuth dependence is here
measured on the deployed set rather than inferred from the pupil geometry. Each mask serves a whole sector, so it is re-checked at
both edges of its own sector. No design loses more than $0.017$\,bit relative to its design axis.

\FloatBarrier
Whether an information-optimal router still sorts into RGB-like channels is
settled by the spectra of the designs above. Figure~S8 shows the four absolute
site responses of the six designed masks at their own field points. Each is
broadband, all four overlap, and their ordering in wavelength shifts from one
design to the next. The optimizer buys conditioning of $\Rmat_\varepsilon$ and
photons rather than band resemblance. Sharp spectral
sorting requires strongly resonant, therefore
angle-sensitive, scattering, and the $\mathrm{NA}\,0.3$ ensemble averages
resonance positions over $\pm11.9^\circ$ inside the oxide ($\pm17.5^\circ$ in
air). It is the same smoothing that flattens the plane-wave spectra of
Fig.~S5. What such a spectrum is worth is set by its throughput and its
purity, which Section~\ref{sec:performance} evaluates.

\subsection{Crossover with the CFA}
\label{sec:performance}
\begin{figure}[tbp]\centering
\includegraphics[width=\linewidth]{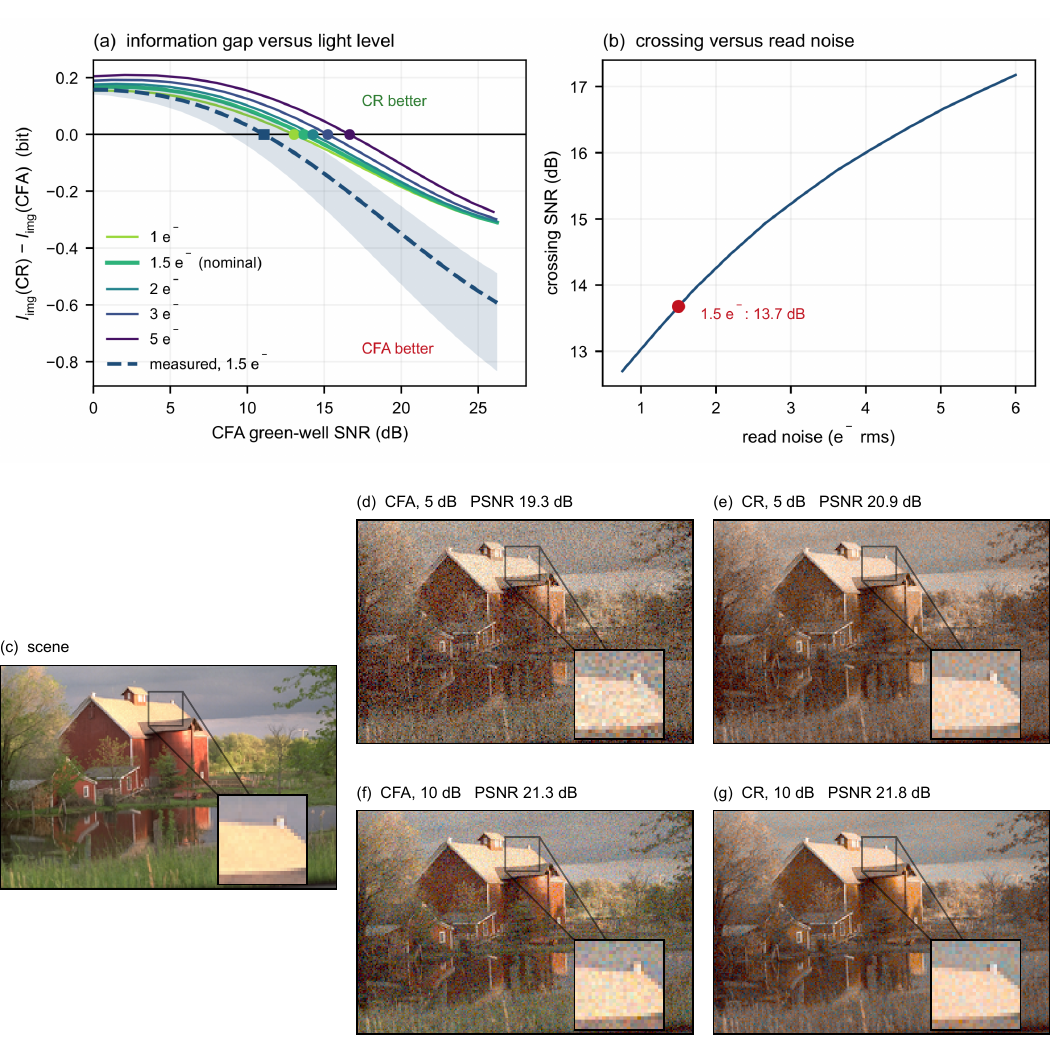}
\caption{\textbf{The CR--CFA crossover and the gap seen as image noise.}
\textbf{(a)}~The imaging-information gap
$\Iimg(\mathrm{CR})-\Iimg(\mathrm{CFA})$
against green-site signal-to-noise ratio, one curve per detector read noise from
$1.0$ to $5.0$~e$^{-}$ rms (light to dark). Each curve is drawn on its own SNR
axis and its filled marker is the zero crossing; the CR leads where the gap is
positive. The dashed curve recomputes the same gap from the fabricated router's
measured routing spectra, with the band around it the scattered-light bracket of
SI Sec.~S8.
\textbf{(b)}~The zero crossing against read noise.
\textbf{(c)}~The scene, carried through the same spectral model as the
reconstructions. \textbf{(d,e)}~Color recovered at $5$\,dB and \textbf{(f,g)}~at
$10$\,dB by the CFA and by the CR, each through its own minimum-mean-square
decoder, for one representative Poisson-and-read-noise draw at that exposure.
This is a colorimetric demonstration: the shared optical blur and demosaicing
are omitted.}
\label{fig:crossover}
\end{figure}

Both imaging systems are evaluated using mutual imaging information $\Iimg$, the same
metric optimized during design. The global performance is computed as an
area-weighted average across the three chief-ray-angle rings (with a ratio of
$1\!:\!8\!:\!16$ from the innermost to outermost ring) to reflect their
respective sensor areas (Fig.~\ref{fig:sector_designs}; Fig.~S3). At the
on-axis position, the optimized color router reaches $\Iimg=1.63$~bits
compared to $1.85$~bits for the matched CFA under nominal exposure, leaving a
deficit of $0.22$~bits. However, this performance order reverses under
low-light conditions because the two sensors are limited by different physical
bounds. Unlike the absorptive CFA, the color router sorts photons and collects
$2.8\times$ more photoelectrons on a quantum-efficiency-weighted basis. While
the CFA's dye absorption provides constant color purity independent of light
level, the color router's throughput advantage dominates in photon-starved
regimes. Plotting the information gap against the light level reveals a
distinct performance crossover (Fig.~\ref{fig:crossover}a).

\begin{figure}[tbp]\centering
\includegraphics[width=\linewidth]{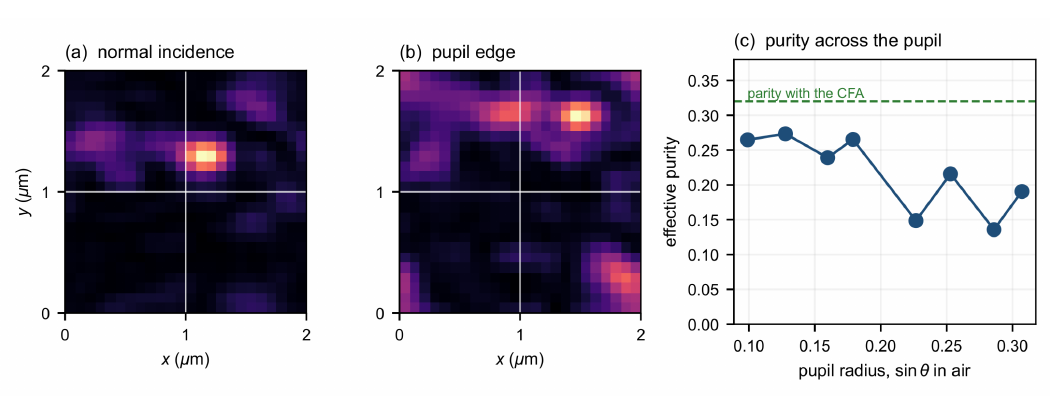}
\caption{\textbf{Walk-off and purity under the imaging-system pupil.}
\textbf{(a,b)}~Detector-plane intensity of the selected mask at $635$\,nm, at
normal incidence and at the edge of the $\mathrm{NA}=0.30$ pupil, with the site
boundaries drawn in white. The sorted lobe is displaced across a boundary while
its power is unchanged. \textbf{(c)}~Effective purity of each quadrature ray
against its pupil radius. The dashed line is the purity at which a CR of
this throughput matches the CFA.}
\label{fig:purity}
\end{figure}

Specifically, below a green-site SNR of $13.7$~dB (corresponding to a mean
charge of $25$~e$^-$ per green site, or one-fifth of the nominal exposure), the
CR preserves more image information than the CFA, leading by up to
$0.17$~bits at the lowest light levels (Fig.~\ref{fig:crossover}a).
Conversely, the CFA dominates under bright light, with the gap widening to
$0.31$~bits at three times nominal exposure where noise is negligible.
Reconstructed color frames below the crossover illustrate this information gap
visually (Fig.~\ref{fig:crossover}c--g): the color router achieves a peak SNR
of $21.8$~dB (vs.\ $21.3$~dB for the CFA) at an input SNR of $10$~dB, and
$20.9$~dB (vs.\ $19.3$~dB) at $5$~dB. Additionally, the crossover point scales
with read noise, rising from $12.7$~dB at $0.75$~e$^-$ rms to $17.2$~dB at
$6.0$~e$^-$ rms (Fig.~\ref{fig:crossover}b). Consequently, noisier sensors
expand the illumination range over which the color router outperforms the
CFA. Decomposing these responses into total throughput $T$ and effective
purity $p$ isolates the physical origin of the bright-light deficit. Here $T$
is the summed site response averaged over the nine evaluation wavelengths,
and $p$ is the least-squares weight with which the pupil-averaged site
response is represented as $T[\,p\,\mathbf S+(1-p)\tfrac14\,]$, where
$\mathbf S$ routes each wavelength band (edges at $500$ and $600$\,nm) entirely
to the sites assigned to it, shared equally when several sites carry one
band, and $p$ is maximized over the $81$ site-to-band assignments. The fit is
taken after the pupil average with equal wavelength weights and no
illuminant or quantum-efficiency weighting, and the CFA is fitted the same
way. The color
router achieves a high on-axis throughput of $T=0.96$ (versus $0.33$ for the
CFA), directing almost all incident light to the photodetectors. However, this
light is insufficiently sorted. The pupil-averaged effective purity of the
router collapses to $0.17$, compared to $0.23$ at normal incidence and $0.50$
for the CFA dyes. This degradation stems from lateral walk-off: under a finite
NA of $0.30$, marginal rays cross the router-to-detector gap at steep angles.
The highly angled diffraction orders spread each sorted wavelength into a
$0.7$--$0.9$~$\mu$m spot across the $1.0$~$\mu$m pixel site. Because the
lateral shift used to route colors is only half a pixel, this substantial spot
size severely blurs the spectral sorting. A ray-by-ray evaluation confirms that
effective purity drops steadily from $0.27$ on the innermost pupil radius to
$0.19$ on the outermost, whereas throughput remains flat
(Fig.~\ref{fig:purity}). Since our optimization objective smoothly balances
throughput and purity, parity with the CFA can be quantified. At $T=0.96$, the
color router matches the reference CFA when its pupil-averaged effective purity
reaches $0.32$. This threshold remains nearly constant (within $0.001$) across
different site assignments. This benchmark rises off-axis where throughput
degrades and purity drops, widening the gap further. Because the design's
normal-incidence purity ($0.23$) starts below this threshold, achieving
bright-light parity requires both enhancing the intrinsic normal-incidence
routing purity and preserving it through the converging imaging-system pupil.

\subsection{Experimental validation}
\label{sec:experiment}

\begin{figure}[tbp]\centering
\includegraphics[width=\linewidth]{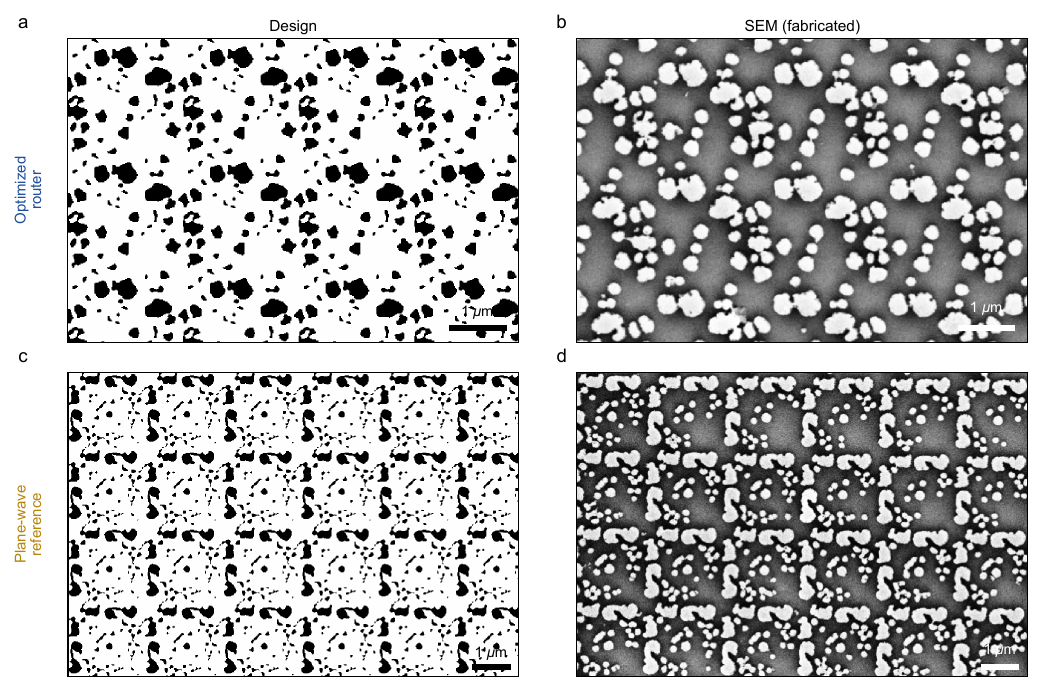}\\[3pt]
\includegraphics[width=\linewidth]{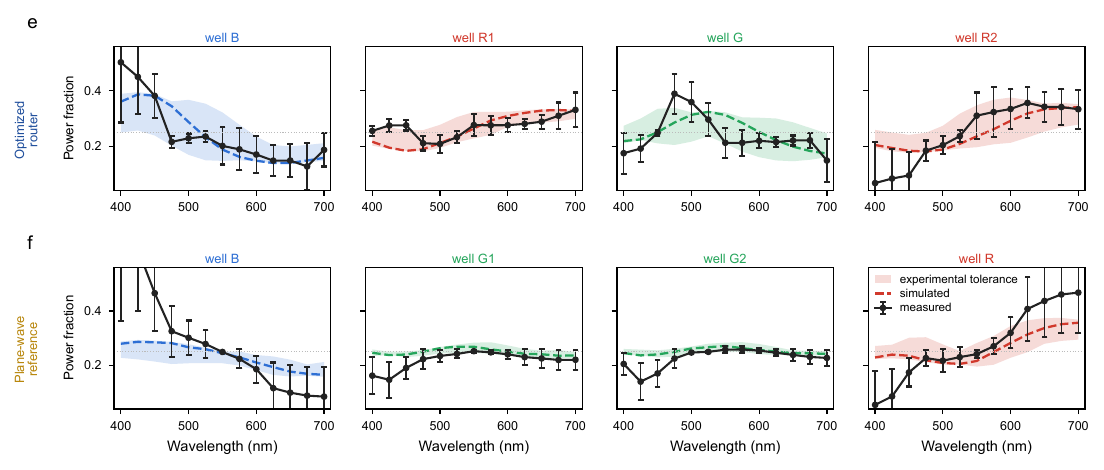}
\caption{\textbf{Fabricated devices under imaging-system-pupil illumination.}
\textbf{(a--d)}~Design layout (SiN black, air white) and top-view scanning
electron micrograph of the fabricated device, for the optimized router
(\textbf{a,b}) and the plane-wave reference (\textbf{c,d}).
\textbf{(e,f)}~Power fraction reaching each of the four detector sites
against wavelength, for the two devices. Points are measured over the whole
illuminated spot; the bars combine the spread across six angular sectors of that
spot, a one-pixel alignment error and the scattered-light correction (SI
Sec.~S8). Dashed curves are simulated; the shaded band spans the
tolerances of SI Secs.~S7 and~S8. Well labels give
the band each site predominantly collects.}
\label{fig:experiment}
\end{figure}

To validate our pupil-averaged forward model, we fabricated both the optimized
router and the plane-wave reference in silicon nitride (SiN). We then
characterized their spectral response at thirteen wavelengths from $400$ to
$700$\,nm under realistic imaging-system-pupil illumination, using a high-NA ($0.90$)
collection objective (Fig.~S10). Scanning electron microscopy (SEM) reveals
clean etches with near-vertical sidewalls for both structures (Fig.~S11).
Comparing the design layouts directly to the fabricated films highlights the
impact of geometric scale: the plane-wave reference is reproduced with high
fidelity, whereas the optimized router, which features more intricate
structures, loses features below $90$\,nm due to lithographic merging and
dropouts (Fig.~\ref{fig:experiment}a--d). This sub-$90$\,nm structural
deviation is further quantified in the spatial design-to-fabrication overlay
of Fig.~S13. We calculated the power allocated to the four detector sites by
averaging the measured field over several hundred supercell periods
(Fig.~S12; Sec.~S8). The experimental allocation matches the simulated forward
model well across the visible spectrum, with minor deviations occurring near
the band edges where routing performance naturally degrades
(Fig.~\ref{fig:experiment}e,f). Evaluating the measured routing spectra through
our information-theoretic framework successfully reproduces the predicted
performance crossover. At nominal exposure, the fabricated device preserves
$\Iimg=1.44^{+0.10}_{-0.19}$~bits of information. It overtakes the CFA below an
input SNR of $11.1^{+2.1}_{-2.3}$~dB, closely matching the simulated prediction
of $13.7$~dB (Fig.~\ref{fig:crossover}a, dashed). The experimental error bars
represent the uncertainty in removing background scattered light: retaining the
entire pedestal shifts the crossover to $8.8$~dB, while subtracting it down to
the cell floor aligns the crossover at $13.2$~dB (Sec.~S8).

\section{Discussion}
\label{sec:discussion}

Replacing the conventional CFA commits the imaging system to a computational
front-end. Because broadband, free-form spectral responses do not suit the
conventional narrowband white-balance/color-correction/demosaic chain
\cite{malvar2004,finlayson2015}, we utilize a simple $3\times4$ linear
demixing operator solved analytically inside our design loop. This approach
introduces a per-region linear decoder and a static site-permutation table.
Both components are fixed at design time and are analogous to existing
lens-shading calibration data \cite{vaillant2008,catrysse2000qe}. The
performance crossover defines the operational regimes of both devices. Below
the crossover, the color router dominates by capturing the photons that a CFA
otherwise absorbs, translating this throughput advantage into superior image
information. Above the crossover, the CFA maintains its lead because the
router is limited by how little of the sorted light reaches the correct site,
rather than by how much it collects, a structural shortfall that the
converging pupil cone deepens but does not create. Three key modeling choices
bound our findings. First, our illumination model assumes a rigidly shifted
circular pupil of uniform radiance under a $\cos^4\theta_{\mathrm{CRA}}$
relative illumination. Second, our electromagnetic solver simulates isolated
periodic cells, meaning that interpixel crosstalk and the behavior at the mask
discontinuities between regions lie outside its scope
\cite{lee2024,anzagira2015}. Third, our comparison is purely colorimetric: the
score represents the zero-spatial-frequency limit, so the shared lens
modulation transfer function (MTF) and the edge color the router adds are not
represented. Scoring by imaging information shifts the design lever from
plane-wave RGB routing efficiency to two quantities measured after
reconstruction under the imaging-system-pupil model: the low-light crossover threshold
and the pupil-averaged routing purity. For our reference imaging-system model, reaching
bright-light parity with the CFA requires raising the pupil-averaged purity
from $0.17$ to $0.32$ at a throughput of $0.96$.

\section{Conclusion and outlook}
\label{sec:conclusion}

In conclusion, this work exposes the fundamental mismatch between the
conventional normal-plane-wave optimization of nanophotonic color routers and
the actual exit pupil illumination geometry, comprising a finite NA and
field-dependent CRA variations, found in real-world imaging systems. Rather
than striving to mimic idealized, narrowband band-pass spectra, we introduced
an end-to-end inverse-design methodology that analytically marginalizes the
optimal linear decoder to directly maximize the mutual imaging information
($\Iimg$) of the reconstructed image. Leveraging the $D_4$ symmetry of the
square pixel lattice, we demonstrated a practical, large-area deployment
strategy capable of tiling $48$ distinct sensor-field regions using only six
unique lithographic mask patterns. The single-layer SiN color router is
predicted to collect $2.8\times$ more photoelectrons than a standard CFA, and
the measured routing fractions of the fabricated device, combined with the
modeled throughput, reproduce the predicted low-light crossover (simulated
$13.7$~dB; measured-routing-based $11.1^{+2.1}_{-2.3}$~dB), marking, to our
knowledge, the first experimental demonstration, from measured routing and a
modeled throughput, of a single-layer color router achieving a low-light
performance crossover against the CFA.

While the CFA maintains its lead in bright-light regimes due to its superior
spectral purity, closing this performance gap requires elevating the
pupil-averaged effective purity of the color router from the current $0.17$ to
the parity threshold of $0.32$. This remaining objective points toward
promising future research directions, including the utilization of taller or
multi-layer nanostructures, the integration of hybrid dye-plus-router stacks,
and the development of richer computational reconstruction pipelines.
Ultimately, our findings establish that the next generation of ultra-sensitive
optical sensors must move beyond isolated device-level efficiency toward
system-level hardware-software co-design. In this new regime, metasurface
color-sensing devices should be optimized and evaluated against the imaging
information preserved under physical imaging-system-pupil illumination, rather
than routing contrast under idealized plane waves.

\section{Materials and methods}
\label{sec:methods}

\subsection{Imaging-system-pupil illumination ensemble}
Each sensor position is modeled as illuminated by a converging, spatially
incoherent finite-numerical-aperture cone whose axis tilts with field position by
the chief-ray angle. The finite-$f$-number illumination is represented as an
incoherent angular sum of plane waves over the exit pupil. The vectorial
Debye--Wolf field of one scene point focused at transverse position
$\mathbf x_0$ is the coherent superposition of pupil plane waves
\cite{wolf1959,richards1959,novotny2012},
\begin{equation}
\bm E_{\mathbf x_0}(\rperp)=\sum_n \sqrt{w_n}\,\widetilde{\bm E}_n(\rperp)\,
e^{-i\kperp^{(n)}\cdot\mathbf x_0},
\qquad
w_n=A(\theta_n,\phi_n)\,\frac{\Delta A_n}{\cos\theta_n},
\label{eq:focus}
\end{equation}
where $\widetilde{\bm E}_n$ is the calibrated detector-plane field of
Eq.~\eqref{eq:norm}, $\Delta A_n$ the direction-cosine area element, $\theta_n$
the polar angle referred to the sensor normal, and $\kperp^{(n)}$ the transverse
wavevector. Because $\Delta A_n/\cos\theta_n$ is the solid-angle element,
Eq.~\eqref{eq:focus} describes a pupil of uniform radiance, a modeling
choice that differs from the aplanatic $\sqrt{\cos\theta}$ apodization of a
sine-condition Richards--Wolf pupil \cite{richards1959,novotny2012} by
$\cos^2\theta$ in intensity across the disk, and at an oblique field point the
difference becomes an asymmetry across the tilted disk. Pupil weighting and the
rigid circular shape assumed off axis both belong to the imaging-system model, whose
scope is set out in Sec.~S1. For a scene incoherent
and uniform over a patch large compared with the van Cittert--Zernike coherence
radius, the off-diagonal phase terms average out and the detected site powers
reduce to the pupil quadrature of Eq.~\eqref{eq:incoh} with solid-angle weights.
The chief-ray tilt displaces every pupil sample on the disk as in
Eq.~\eqref{eq:tiltshift}, so the source is a function of two field coordinates,
the tilt magnitude $\theta_{\mathrm{CRA}}$ and the tilt azimuth
$\phi_{\mathrm{CRA}}$. All pupil directions, the disk radius and
$\theta_{\mathrm{CRA}}$ are specified on the air side. Optimization uses an
eight-ray, two-polarization reduction
of the shifted-disk quadrature, the two orthogonal polarizations averaged with
equal weight to the unpolarized response, and every reported value uses the
corresponding sixteen-ray rule. Convergence of the reduced quadrature is verified by
quadrature refinement in Sec.~S1 (Fig.~S1)
\cite{richards1959,youngworth2000,novotny2012}. Pupil weights are normalized to a
$\cos^4\theta_{\mathrm{CRA}}$ relative-illumination model used as a reference
lens model, since CRA-matched modules with shifted microlenses commonly fall off
more slowly \cite{catrysse2000qe,vaillant2008,huo2010}; measured or ray-traced lens-specific pupil weights can be substituted without changing the formulation.

\subsection{Electromagnetic forward model}
Each pupil plane wave is one rigorous coupled-wave analysis solve
\cite{moharam1995a} in the differentiable GPU implementation TORCWA
\cite{kim2023}, on a single free-form binary SiN layer of height $600$\,nm on a
$2\,\mu$m four-site cell, air above and oxide below with the detector plane in the
oxide. Dispersion uses measured optical constants for SiN \cite{luke2015} and
SiO$_2$ \cite{malitson1965}. Site energies are calibrated to the
cell-total transmitted power: with $A_{\mathrm{cell}}=L_xL_y$ the cell area,
$\bm E^{\mathrm{raw}}_n$ the simulated detector-plane field and
$I_{b,n}=\int_{\Omega_b}\|\bm E^{\mathrm{raw}}_n\|^2dA$ its energy in site $b$,
\begin{equation}
\widetilde{\bm E}_n=a_n\bm E^{\mathrm{raw}}_n,
\qquad
a_n^2=\frac{A_{\mathrm{cell}}T_n}{\sum_{c=1}^{4}I_{c,n}},
\label{eq:norm}
\end{equation}
with $T_n$ the longitudinal-power-normalized total transmittance. Detector-plane conventions,
order truncation, the $(8,8)$ differentiable surrogate and the hard-mask
evaluation settings are specified in Sec.~S2.

\subsection{Imaging-information objective and decoder}
Site responses are converted to expected photoelectrons through the photon
quadrature: with $L_E$ the relative spectral power distribution of the
illuminant and $Q$ a common representative back-side-illuminated silicon quantum
efficiency \cite{vereecke2015},
\begin{equation}
\Phi_i=\frac{L_E(\lambda_i)\,\Delta\lambda_i\,\lambda_i}
{\sum_j L_E(\lambda_j)\,\Delta\lambda_j\,\lambda_j},
\qquad
R_{b,i}=\overline P_b(\lambda_i)\,\Phi_i\,Q(\lambda_i),
\label{eq:electron}
\end{equation}
giving the four-site electron operator $\Rmat_\varepsilon$. Retaining the
wavelength factor before the photon-defined $Q$ is what makes
$\Rmat_\varepsilon$ an electron operator, and the common constants are absorbed
into the exposure scale. The four-site signal carries a signal-dependent
shot-plus-read covariance. The design objective is the imaging information
$\Iimg\equiv\tfrac14 I(\mathbf Y;\mathbf Z)$, a quarter of the mutual information
between the four noisy raw measurements and the CIE XYZ image \cite{cie0152018}, with the
closed form of Eq.~\eqref{eq:iimg} in terms of the generalized eigenvalues of the
posterior color covariance against the reflectance-prior covariance
\cite{babelcolor2012}. The optimal linear decoder that minimizes color error is
solved in closed form and re-solved at every iterate, so the geometry is
differentiated through the decoder rather than against a trained one. $\Iimg$ is
invariant under invertible linear maps of the four measurements and is reported in
bits per raw pixel, averaged over the exposure ladder after the logarithm rather
than by averaging posteriors. The full
cross-spectral-density derivation and the information model are given in
Secs.~S1 and~S3.

\subsection{Inverse design and deployment}
\begin{figure}[tbp]
\centering
\includegraphics[width=\linewidth]{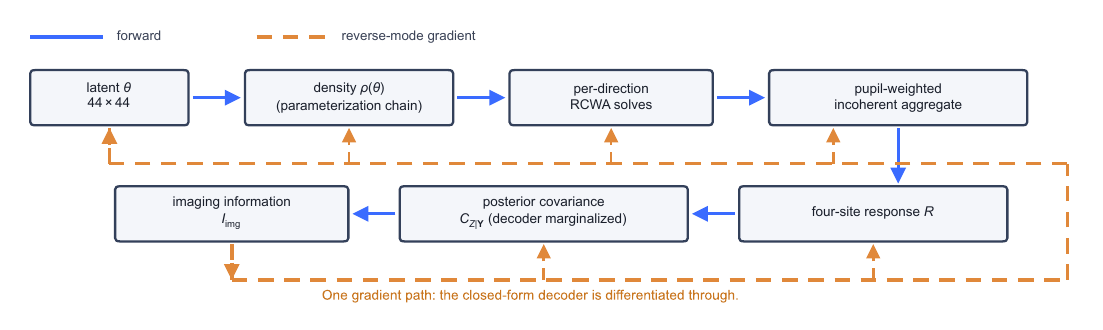}
\caption{\textbf{Forward and backward path of the inverse design.}
The forward pass maps the latent field through the parameterization chain,
per-direction RCWA solves, and pupil-weighted aggregation to the four-site
response; marginalizing the closed-form decoder gives the posterior color
covariance, whose log-determinant is the imaging information $\Iimg$.
The reverse-mode gradient (dashed) returns through the entire chain,
including the decoder, to update the latent.}
\label{fig:gradient_path}
\end{figure}

The pattern is parameterized as a $44\times44$ latent field upscaled to a
$128\times128$ raster through a conic filter of radius $45$\,nm and a tanh
projection in the robust erosion/dilation formulation
\cite{sigmund2007,wang2011}, giving $1936$ free parameters. The imaging
information is maximized by reverse-mode gradient descent through the pupil
average and the electromagnetic solver; Figure~\ref{fig:gradient_path} traces
the forward and backward pass of this loop. Field-dependent deployment tiles the
two-dimensional $(\theta_{\mathrm{CRA}},\phi)$ domain into rings in chief-ray
angle crossed with azimuth sectors, and the $D_4$ symmetry of the square
four-site lattice reduces the $48$ deployed regions to six designed masks. The
tiling itself adds no process step, since all $48$ regions share one
deposition, one etch and one height and differ only in a lithography pattern
that is position-dependent in any case. Fabrication-error sensitivity is
specified in Sec.~S7, and optimization schedules and the noise model in Secs.~S3 and~S6.

\FloatBarrier
\subsection*{Code availability}
A reference implementation of the core method, the imaging-system-pupil forward
model and the imaging-information objective, is available at
\url{https://github.com/hyoseokp/information-optimal-color-router} and archived
on Zenodo at \url{https://doi.org/10.5281/zenodo.22099957}.

\subsection*{Data availability}
The final design masks (the six per-slot designs r1a--r3c and the champion)
with their manifest of slots, seeds and
angles, together with the derived response caches for the simulation figures,
are available in the public repository at
\url{https://github.com/hyoseokp/information-optimal-color-router}, in its
data/ directory, and archived on Zenodo at
\url{https://doi.org/10.5281/zenodo.22099957}. The plane-wave reference design, the experimental measurements, the electron micrographs and the
SEM-derived masks, and the numerical source data for the remaining figures are
available from the corresponding author on request.

\section*{Acknowledgements}
This work was supported by the BK21 FOUR Program, the Chungnam National
University Research Grant 2026, the Basic Science Research Program through the
National Research Foundation of Korea (NRF) funded by the Ministry of Education
(RS-2020-NR049597), the creation of the quantum information science R\&D
ecosystem (based on human resources) through the National Research Foundation
of Korea (NRF) funded by the Ministry of Science and ICT (RS-2023-00256050),
the Global--Learning \& Academic research institution for Master's, PhD
students, and Postdocs (G-LAMP) Program of the National Research Foundation of
Korea (NRF) grant funded by the Ministry of Education (RS-2025-25442707), the
National Research Foundation of Korea (NRF) grant funded by the Korea
government (Ministry of Science and ICT) (RS-2025-24533176), the Technology
Innovation Program Development Program (RS-2023-00235080) funded by the
Ministry of Trade, Industry \& Energy (MOTIE, Korea), and the National Research
Foundation of Korea (NRF) grant funded by the Korea government (MSIT)
(RS-2026-25606142).

\section*{Conflict of interest}
The authors declare no conflict of interest.

\section*{Author contributions}
H.P. and S.P. contributed equally to this work.
H.P.: methodology, software, simulation, analysis, and writing of the original
draft.
S.P. and M.L.: experimental setup, device fabrication, and measurement.
M.L.: supervision of the fabrication and measurement, and corresponding author.
Y.P.: conceptualization, supervision, writing (review and editing), and
corresponding author.

\section*{Supplementary information}
Supplementary information accompanies the manuscript on the Light: Science \&
Applications website (\url{http://www.nature.com/lsa/}).

\FloatBarrier
\begingroup
\catcode`\&=12\relax
\footnotesize
\setlength{\bibsep}{0pt plus 0.2ex}
\bibliographystyle{naturemag}
\bibliography{refs_lsa}

\begin{thebibliography}{10}
\expandafter\ifx\csname url\endcsname\relax
  \def\url#1{\texttt{#1}}\fi
\expandafter\ifx\csname urlprefix\endcsname\relax\def\urlprefix{URL }\fi
\providecommand{\bibinfo}[2]{#2}
\providecommand{\eprint}[2][]{\url{#2}}

\bibitem{bayer1976}
\bibinfo{author}{Bayer, B.~E.}
\newblock \bibinfo{title}{Color imaging array} (\bibinfo{year}{1976}).
\newblock \urlprefix\url{https://patents.google.com/patent/US3971065A/en}.
\newblock \bibinfo{note}{U.S. Patent 3,971,065, filed 5 March 1975, granted 20
  July 1976, assignee Eastman Kodak Company}.

\bibitem{nishiwaki2013}
\bibinfo{author}{Nishiwaki, S.}, \bibinfo{author}{Nakamura, T.},
  \bibinfo{author}{Hiramoto, M.}, \bibinfo{author}{Fujii, T.} \&
  \bibinfo{author}{Suzuki, M.-a.}
\newblock \bibinfo{title}{Efficient colour splitters for high-pixel-density
  image sensors}.
\newblock \emph{\bibinfo{journal}{Nat. Photonics}}
  \textbf{\bibinfo{volume}{7}}, \bibinfo{pages}{240--246}
  (\bibinfo{year}{2013}).

\bibitem{miyata2021}
\bibinfo{author}{Miyata, M.}, \bibinfo{author}{Nemoto, N.},
  \bibinfo{author}{Shikama, K.}, \bibinfo{author}{Kobayashi, F.} \&
  \bibinfo{author}{Hashimoto, T.}
\newblock \bibinfo{title}{Full-color-sorting metalenses for high-sensitivity
  image sensors}.
\newblock \emph{\bibinfo{journal}{Optica}} \textbf{\bibinfo{volume}{8}},
  \bibinfo{pages}{1596--1604} (\bibinfo{year}{2021}).

\bibitem{zou2022}
\bibinfo{author}{Zou, X.} \emph{et~al.}
\newblock \bibinfo{title}{Pixel-level {B}ayer-type colour router based on
  metasurfaces}.
\newblock \emph{\bibinfo{journal}{Nat. Commun.}} \textbf{\bibinfo{volume}{13}},
  \bibinfo{pages}{3288} (\bibinfo{year}{2022}).

\bibitem{lee2024}
\bibinfo{author}{Lee, S.} \emph{et~al.}
\newblock \bibinfo{title}{Inverse design of color routers in {CMOS} image
  sensors: toward minimizing interpixel crosstalk}.
\newblock \emph{\bibinfo{journal}{Nanophotonics}}
  \textbf{\bibinfo{volume}{13}}, \bibinfo{pages}{3895--3914}
  (\bibinfo{year}{2024}).

\bibitem{rao2025}
\bibinfo{author}{Rao, R.}, \bibinfo{author}{Shi, Y.}, \bibinfo{author}{Wang,
  Z.}, \bibinfo{author}{Wan, S.} \& \bibinfo{author}{Li, Z.}
\newblock \bibinfo{title}{On-chip cascaded metasurfaces for visible wavelength
  division multiplexing and color-routing meta-display}.
\newblock \emph{\bibinfo{journal}{Nano Lett.}} \textbf{\bibinfo{volume}{25}},
  \bibinfo{pages}{2452--2458} (\bibinfo{year}{2025}).

\bibitem{kim2024router}
\bibinfo{author}{Kim, C.} \emph{et~al.}
\newblock \bibinfo{title}{Freeform metasurface color router for deep submicron
  pixel image sensors}.
\newblock \emph{\bibinfo{journal}{Sci. Adv.}} \textbf{\bibinfo{volume}{10}},
  \bibinfo{pages}{eadn9000} (\bibinfo{year}{2024}).

\bibitem{li2022bayer}
\bibinfo{author}{Li, J.} \emph{et~al.}
\newblock \bibinfo{title}{Single-layer {Bayer} metasurface via inverse design}.
\newblock \emph{\bibinfo{journal}{ACS Photonics}} \textbf{\bibinfo{volume}{9}},
  \bibinfo{pages}{2607--2613} (\bibinfo{year}{2022}).

\bibitem{park2026ropp}
\bibinfo{author}{Park, H.}, \bibinfo{author}{Kim, S.}, \bibinfo{author}{Choi,
  D.-Y.}, \bibinfo{author}{Park, Y.} \& \bibinfo{author}{Lee, M.}
\newblock \bibinfo{title}{Metasurface color routers inverse-designed with a
  deep-learning-accelerated genetic algorithm for high-efficiency imaging}.
\newblock \emph{\bibinfo{journal}{Rep. Prog. Phys.}}  (\bibinfo{year}{2026}).
\newblock \bibinfo{note}{\url{https://doi.org/10.1088/1361-6633/ae9e8c}}.

\bibitem{go2025iedm}
\bibinfo{author}{Go, J.} \emph{et~al.}
\newblock \bibinfo{title}{A 2-layer, 0.7\,$\mu$m-pitch dual photodiode pixel
  {CMOS} image sensor with metaphotonic color router}.
\newblock In \emph{\bibinfo{booktitle}{2025 IEEE International Electron Devices
  Meeting (IEDM)}}, \bibinfo{pages}{1--4} (\bibinfo{year}{2025}).

\bibitem{yu2014}
\bibinfo{author}{Yu, N.} \& \bibinfo{author}{Capasso, F.}
\newblock \bibinfo{title}{Flat optics with designer metasurfaces}.
\newblock \emph{\bibinfo{journal}{Nat. Mater.}} \textbf{\bibinfo{volume}{13}},
  \bibinfo{pages}{139--150} (\bibinfo{year}{2014}).

\bibitem{fossum1997}
\bibinfo{author}{Fossum, E.~R.}
\newblock \bibinfo{title}{{CMOS} image sensors: electronic camera-on-a-chip}.
\newblock \emph{\bibinfo{journal}{IEEE Trans. Electron Devices}}
  \textbf{\bibinfo{volume}{44}}, \bibinfo{pages}{1689--1698}
  (\bibinfo{year}{1997}).

\bibitem{fossum2014}
\bibinfo{author}{Fossum, E.~R.} \& \bibinfo{author}{Hondongwa, D.~B.}
\newblock \bibinfo{title}{A review of the pinned photodiode for {CCD} and
  {CMOS} image sensors}.
\newblock \emph{\bibinfo{journal}{IEEE J. Electron Devices Soc.}}
  \textbf{\bibinfo{volume}{2}}, \bibinfo{pages}{33--43} (\bibinfo{year}{2014}).

\bibitem{hasinoff2010}
\bibinfo{author}{Hasinoff, S.~W.}, \bibinfo{author}{Durand, F.} \&
  \bibinfo{author}{Freeman, W.~T.}
\newblock \bibinfo{title}{Noise-optimal capture for high dynamic range
  photography}.
\newblock In \emph{\bibinfo{booktitle}{2010 {IEEE} Computer Society Conference
  on Computer Vision and Pattern Recognition}}, \bibinfo{pages}{553--560}
  (\bibinfo{year}{2010}).

\bibitem{vaillant2004}
\bibinfo{author}{Vaillant, J.} \& \bibinfo{author}{Hirigoyen, F.}
\newblock \bibinfo{title}{Optical simulation for {CMOS} imager microlens
  optimization}.
\newblock In \emph{\bibinfo{booktitle}{SPIE Proceedings}}, vol.
  \bibinfo{volume}{5459}, \bibinfo{pages}{200} (\bibinfo{year}{2004}).

\bibitem{vaillant2008}
\bibinfo{author}{Vaillant, J.} \emph{et~al.}
\newblock \bibinfo{title}{Versatile method for optical performances
  characterization of off-axis {CMOS} pixels with microlens radial shift}.
\newblock In \emph{\bibinfo{booktitle}{{SPIE} Proceedings}}, vol.
  \bibinfo{volume}{6817}, \bibinfo{pages}{681707} (\bibinfo{year}{2008}).

\bibitem{hong2025vlsi}
\bibinfo{author}{Hong, J.} \emph{et~al.}
\newblock \bibinfo{title}{Adaptive metasurface microlens array for
  ultra-wide-angle {CMOS} image sensors}.
\newblock In \emph{\bibinfo{booktitle}{2025 Symposium on VLSI Technology and
  Circuits}}, \bibinfo{pages}{1--3} (\bibinfo{year}{2025}).

\bibitem{vaillant2007}
\bibinfo{author}{Vaillant, J.}, \bibinfo{author}{Crocherie, A.},
  \bibinfo{author}{Hirigoyen, F.}, \bibinfo{author}{Cadien, A.} \&
  \bibinfo{author}{Pond, J.}
\newblock \bibinfo{title}{Uniform illumination and rigorous electromagnetic
  simulations applied to {CMOS} image sensors}.
\newblock \emph{\bibinfo{journal}{Opt. Express}} \textbf{\bibinfo{volume}{15}},
  \bibinfo{pages}{5494--5503} (\bibinfo{year}{2007}).

\bibitem{zhao2021}
\bibinfo{author}{Zhao, N.}, \bibinfo{author}{Catrysse, P.~B.} \&
  \bibinfo{author}{Fan, S.}
\newblock \bibinfo{title}{Perfect {RGB}-{IR} color routers for sub-wavelength
  size {CMOS} image sensor pixels}.
\newblock \emph{\bibinfo{journal}{Adv. Photonics Res.}}
  \textbf{\bibinfo{volume}{2}}, \bibinfo{pages}{2000048}
  (\bibinfo{year}{2021}).

\bibitem{jeon2025}
\bibinfo{author}{Jeon, J.}, \bibinfo{author}{Park, C.}, \bibinfo{author}{Heo,
  D.}, \bibinfo{author}{Chung, H.} \& \bibinfo{author}{Jang, M.~S.}
\newblock \bibinfo{title}{Inverse design of nanophotonic color router robust to
  oblique incidence}.
\newblock \emph{\bibinfo{journal}{Adv. Opt. Mater.}}
  \textbf{\bibinfo{volume}{14}}, \bibinfo{pages}{e01697}
  (\bibinfo{year}{2026}).

\bibitem{kim2026stitch}
\bibinfo{author}{Kim, D.}, \bibinfo{author}{Han, J.}, \bibinfo{author}{Lee,
  S.}, \bibinfo{author}{Jang, M.~S.} \& \bibinfo{author}{Chung, H.}
\newblock \bibinfo{title}{Suppressing stitching errors in full-sensor-plane
  color routers via optical structural similarity}.
\newblock \emph{\bibinfo{journal}{Nanophotonics}}
  \textbf{\bibinfo{volume}{15}}, \bibinfo{pages}{e70220}
  (\bibinfo{year}{2026}).

\bibitem{seo2026}
\bibinfo{author}{Seo, D.}, \bibinfo{author}{Um, S.}, \bibinfo{author}{Lee, S.},
  \bibinfo{author}{Ye, J.~C.} \& \bibinfo{author}{Chung, H.}
\newblock \bibinfo{title}{Physics-guided and fabrication-aware inverse design
  of photonic devices using diffusion models}.
\newblock \emph{\bibinfo{journal}{{ACS} Photonics}}
  \textbf{\bibinfo{volume}{13}}, \bibinfo{pages}{363--372}
  (\bibinfo{year}{2026}).

\bibitem{miyata2026}
\bibinfo{author}{Miyata, M.}, \bibinfo{author}{Shikama, K.},
  \bibinfo{author}{Takehara, H.}, \bibinfo{author}{Ohta, J.} \&
  \bibinfo{author}{Hashimoto, T.}
\newblock \bibinfo{title}{High-sensitivity {RGB}--{NIR} image sensor with
  dispersion-engineered meta-optics}.
\newblock \emph{\bibinfo{journal}{ACS Photonics}}
  \textbf{\bibinfo{volume}{13}}, \bibinfo{pages}{849--860}
  (\bibinfo{year}{2026}).

\bibitem{johlin2021}
\bibinfo{author}{Johlin, E.}
\newblock \bibinfo{title}{Nanophotonic color splitters for high-efficiency
  imaging}.
\newblock \emph{\bibinfo{journal}{iScience}} \textbf{\bibinfo{volume}{24}},
  \bibinfo{pages}{102268} (\bibinfo{year}{2021}).

\bibitem{lin2021e2e}
\bibinfo{author}{Lin, Z.} \emph{et~al.}
\newblock \bibinfo{title}{End-to-end nanophotonic inverse design for imaging
  and polarimetry}.
\newblock \emph{\bibinfo{journal}{Nanophotonics}}
  \textbf{\bibinfo{volume}{10}}, \bibinfo{pages}{1177--1187}
  (\bibinfo{year}{2021}).

\bibitem{sun2025}
\bibinfo{author}{Sun, J.} \emph{et~al.}
\newblock \bibinfo{title}{Collaborative on-sensor array cameras}.
\newblock \emph{\bibinfo{journal}{ACM Trans. Graph.}}
  \textbf{\bibinfo{volume}{44}}, \bibinfo{pages}{1--18} (\bibinfo{year}{2025}).

\bibitem{vora1993}
\bibinfo{author}{Vora, P.~L.} \& \bibinfo{author}{Trussell, H.~J.}
\newblock \bibinfo{title}{Measure of goodness of a set of color-scanning
  filters}.
\newblock \emph{\bibinfo{journal}{J. Opt. Soc. Am. A}}
  \textbf{\bibinfo{volume}{10}}, \bibinfo{pages}{1499--1508}
  (\bibinfo{year}{1993}).

\bibitem{hirakawa2008}
\bibinfo{author}{Hirakawa, K.} \& \bibinfo{author}{Wolfe, P.}
\newblock \bibinfo{title}{Spatio-spectral color filter array design for optimal
  image recovery}.
\newblock \emph{\bibinfo{journal}{IEEE Trans. Image Process.}}
  \textbf{\bibinfo{volume}{17}}, \bibinfo{pages}{1876--1890}
  (\bibinfo{year}{2008}).

\bibitem{pinkard2025}
\bibinfo{author}{Pinkard, H.} \emph{et~al.}
\newblock \bibinfo{title}{Information-driven design of imaging systems}.
\newblock In \emph{\bibinfo{booktitle}{Advances in Neural Information
  Processing Systems}} (\bibinfo{year}{2025}).

\bibitem{kabuli2026}
\bibinfo{author}{Kabuli, L.~A.}, \bibinfo{author}{Pinkard, H.},
  \bibinfo{author}{Markley, E.}, \bibinfo{author}{Hung, C.~S.} \&
  \bibinfo{author}{Waller, L.}
\newblock \bibinfo{title}{Designing lensless imaging systems to maximize
  information capture}.
\newblock \emph{\bibinfo{journal}{Optica}} \textbf{\bibinfo{volume}{13}},
  \bibinfo{pages}{227--235} (\bibinfo{year}{2026}).

\bibitem{wolf1959}
\bibinfo{author}{Wolf, E.}
\newblock \bibinfo{title}{Electromagnetic diffraction in optical systems - {I.
  An} integral representation of the image field}.
\newblock \emph{\bibinfo{journal}{Proc. R. Soc. Lond. A}}
  \textbf{\bibinfo{volume}{253}}, \bibinfo{pages}{349--357}
  (\bibinfo{year}{1959}).

\bibitem{richards1959}
\bibinfo{author}{Richards, B.} \& \bibinfo{author}{Wolf, E.}
\newblock \bibinfo{title}{Electromagnetic diffraction in optical systems, {II}.
  structure of the image field in an aplanatic system}.
\newblock \emph{\bibinfo{journal}{Proc. R. Soc. Lond. A}}
  \textbf{\bibinfo{volume}{253}}, \bibinfo{pages}{358--379}
  (\bibinfo{year}{1959}).

\bibitem{novotny2012}
\bibinfo{author}{Novotny, L.} \& \bibinfo{author}{Hecht, B.}
\newblock \emph{\bibinfo{title}{Principles of Nano-Optics}}
  (\bibinfo{publisher}{Cambridge University Press}, \bibinfo{year}{2012}),
  \bibinfo{edition}{2nd} edn.

\bibitem{vancittert1934}
\bibinfo{author}{van Cittert, P.}
\newblock \bibinfo{title}{Die wahrscheinliche {Schwingungsverteilung} in einer
  von einer {Lichtquelle} direkt oder mittels einer {Linse} beleuchteten
  {Ebene}}.
\newblock \emph{\bibinfo{journal}{Physica}} \textbf{\bibinfo{volume}{1}},
  \bibinfo{pages}{201--210} (\bibinfo{year}{1934}).

\bibitem{zernike1938}
\bibinfo{author}{Zernike, F.}
\newblock \bibinfo{title}{The concept of degree of coherence and its
  application to optical problems}.
\newblock \emph{\bibinfo{journal}{Physica}} \textbf{\bibinfo{volume}{5}},
  \bibinfo{pages}{785--795} (\bibinfo{year}{1938}).

\bibitem{wolf1954}
\bibinfo{author}{Wolf, E.}
\newblock \bibinfo{title}{A macroscopic theory of interference and diffraction
  of light from finite sources, {I. Fields} with a narrow spectral range}.
\newblock \emph{\bibinfo{journal}{Proc. R. Soc. Lond. A}}
  \textbf{\bibinfo{volume}{225}}, \bibinfo{pages}{96--111}
  (\bibinfo{year}{1954}).

\bibitem{catrysse2000qe}
\bibinfo{author}{Catrysse, P.~B.}, \bibinfo{author}{Liu, X.} \&
  \bibinfo{author}{El~Gamal, A.}
\newblock \bibinfo{title}{{QE} reduction due to pixel vignetting in {CMOS}
  image sensors}.
\newblock In \emph{\bibinfo{booktitle}{SPIE Proceedings}}, vol.
  \bibinfo{volume}{3965}, \bibinfo{pages}{420--430} (\bibinfo{year}{2000}).

\bibitem{huo2010}
\bibinfo{author}{Huo, Y.}, \bibinfo{author}{Fesenmaier, C.~C.} \&
  \bibinfo{author}{Catrysse, P.~B.}
\newblock \bibinfo{title}{Microlens performance limits in sub-2$\mu$m pixel
  {CMOS} image sensors}.
\newblock \emph{\bibinfo{journal}{Opt. Express}} \textbf{\bibinfo{volume}{18}},
  \bibinfo{pages}{5861--5872} (\bibinfo{year}{2010}).

\bibitem{cie0152018}
\bibinfo{author}{{Commission Internationale de l'Eclairage}}.
\newblock \emph{\bibinfo{title}{Colorimetry, 4th Edition}}.
\newblock \bibinfo{organization}{CIE} (\bibinfo{year}{2018}).

\bibitem{vereecke2015}
\bibinfo{author}{Vereecke, B.} \emph{et~al.}
\newblock \bibinfo{title}{Quantum efficiency and dark current evaluation of a
  backside illuminated {CMOS} image sensor}.
\newblock \emph{\bibinfo{journal}{Jpn. J. Appl. Phys.}}
  \textbf{\bibinfo{volume}{54}}, \bibinfo{pages}{04DE09}
  (\bibinfo{year}{2015}).

\bibitem{babelcolor2012}
\bibinfo{author}{{BabelColor}}.
\newblock \bibinfo{title}{{ColorChecker} averaged spectral data from 30
  charts}.
\newblock \bibinfo{howpublished}{CGATS data file
  \texttt{CC\_Avg30\_spectrum\_CGATS.txt}, April 2012} (\bibinfo{year}{2012}).
\newblock \urlprefix\url{https://babelcolor.com/colorchecker-2.htm}.
\newblock \bibinfo{note}{Accessed 5 August 2026}.

\bibitem{catrysse2002}
\bibinfo{author}{Catrysse, P.~B.} \& \bibinfo{author}{Wandell, B.~A.}
\newblock \bibinfo{title}{Optical efficiency of image sensor pixels}.
\newblock \emph{\bibinfo{journal}{J. Opt. Soc. Am. A}}
  \textbf{\bibinfo{volume}{19}}, \bibinfo{pages}{1610--1620}
  (\bibinfo{year}{2002}).

\bibitem{agranov2003}
\bibinfo{author}{Agranov, G.}, \bibinfo{author}{Berezin, V.} \&
  \bibinfo{author}{Tsai, R.}
\newblock \bibinfo{title}{Crosstalk and microlens study in a color {CMOS} image
  sensor}.
\newblock \emph{\bibinfo{journal}{IEEE Trans. Electron Devices}}
  \textbf{\bibinfo{volume}{50}}, \bibinfo{pages}{4--11} (\bibinfo{year}{2003}).

\bibitem{wakabayashi2010}
\bibinfo{author}{Wakabayashi, H.} \emph{et~al.}
\newblock \bibinfo{title}{A 1/2.3-inch 10.3{Mpixel} 50frame/s back-illuminated
  {CMOS} image sensor}.
\newblock In \emph{\bibinfo{booktitle}{2010 {IEEE} International Solid-State
  Circuits Conference - ({ISSCC})}}, \bibinfo{pages}{410--411}
  (\bibinfo{year}{2010}).

\bibitem{moharam1981}
\bibinfo{author}{Moharam, M.~G.} \& \bibinfo{author}{Gaylord, T.~K.}
\newblock \bibinfo{title}{Rigorous coupled-wave analysis of planar-grating
  diffraction}.
\newblock \emph{\bibinfo{journal}{J. Opt. Soc. Am.}}
  \textbf{\bibinfo{volume}{71}}, \bibinfo{pages}{811--818}
  (\bibinfo{year}{1981}).

\bibitem{moharam1995a}
\bibinfo{author}{Moharam, M.~G.}, \bibinfo{author}{Gaylord, T.~K.},
  \bibinfo{author}{Grann, E.~B.} \& \bibinfo{author}{Pommet, D.~A.}
\newblock \bibinfo{title}{Formulation for stable and efficient implementation
  of the rigorous coupled-wave analysis of binary gratings}.
\newblock \emph{\bibinfo{journal}{J. Opt. Soc. Am. A}}
  \textbf{\bibinfo{volume}{12}}, \bibinfo{pages}{1068--1076}
  (\bibinfo{year}{1995}).

\bibitem{moharam1995b}
\bibinfo{author}{Moharam, M.~G.}, \bibinfo{author}{Gaylord, T.~K.},
  \bibinfo{author}{Pommet, D.~A.} \& \bibinfo{author}{Grann, E.~B.}
\newblock \bibinfo{title}{Stable implementation of the rigorous coupled-wave
  analysis for surface-relief gratings: enhanced transmittance matrix
  approach}.
\newblock \emph{\bibinfo{journal}{J. Opt. Soc. Am. A}}
  \textbf{\bibinfo{volume}{12}}, \bibinfo{pages}{1077--1086}
  (\bibinfo{year}{1995}).

\bibitem{li1996a}
\bibinfo{author}{Li, L.}
\newblock \bibinfo{title}{Formulation and comparison of two recursive matrix
  algorithms for modeling layered diffraction gratings}.
\newblock \emph{\bibinfo{journal}{J. Opt. Soc. Am. A}}
  \textbf{\bibinfo{volume}{13}}, \bibinfo{pages}{1024--1035}
  (\bibinfo{year}{1996}).

\bibitem{li1996b}
\bibinfo{author}{Li, L.}
\newblock \bibinfo{title}{Use of {F}ourier series in the analysis of
  discontinuous periodic structures}.
\newblock \emph{\bibinfo{journal}{J. Opt. Soc. Am. A}}
  \textbf{\bibinfo{volume}{13}}, \bibinfo{pages}{1870--1876}
  (\bibinfo{year}{1996}).

\bibitem{kim2023}
\bibinfo{author}{Kim, C.} \& \bibinfo{author}{Lee, B.}
\newblock \bibinfo{title}{{TORCWA}: {GPU}-accelerated {F}ourier modal method
  and gradient-based optimization for metasurface design}.
\newblock \emph{\bibinfo{journal}{Comput. Phys. Commun.}}
  \textbf{\bibinfo{volume}{282}}, \bibinfo{pages}{108552}
  (\bibinfo{year}{2023}).

\bibitem{sigmund2007}
\bibinfo{author}{Sigmund, O.}
\newblock \bibinfo{title}{Morphology-based black and white filters for topology
  optimization}.
\newblock \emph{\bibinfo{journal}{Struct. Multidiscip. Optim.}}
  \textbf{\bibinfo{volume}{33}}, \bibinfo{pages}{401--424}
  (\bibinfo{year}{2007}).

\bibitem{wang2011}
\bibinfo{author}{Wang, F.}, \bibinfo{author}{Lazarov, B.~S.} \&
  \bibinfo{author}{Sigmund, O.}
\newblock \bibinfo{title}{On projection methods, convergence and robust
  formulations in topology optimization}.
\newblock \emph{\bibinfo{journal}{Struct. Multidiscip. Optim.}}
  \textbf{\bibinfo{volume}{43}}, \bibinfo{pages}{767--784}
  (\bibinfo{year}{2011}).

\bibitem{zhou2015}
\bibinfo{author}{Zhou, M.}, \bibinfo{author}{Lazarov, B.~S.},
  \bibinfo{author}{Wang, F.} \& \bibinfo{author}{Sigmund, O.}
\newblock \bibinfo{title}{Minimum length scale in topology optimization by
  geometric constraints}.
\newblock \emph{\bibinfo{journal}{Comput. Methods Appl. Mech. Eng.}}
  \textbf{\bibinfo{volume}{293}}, \bibinfo{pages}{266--282}
  (\bibinfo{year}{2015}).

\bibitem{malvar2004}
\bibinfo{author}{Malvar, H.~S.}, \bibinfo{author}{He, L.-w.} \&
  \bibinfo{author}{Cutler, R.}
\newblock \bibinfo{title}{High-quality linear interpolation for demosaicing of
  {B}ayer-patterned color images}.
\newblock In \emph{\bibinfo{booktitle}{2004 {IEEE} International Conference on
  Acoustics, Speech, and Signal Processing}}, vol.~\bibinfo{volume}{3},
  \bibinfo{pages}{iii--485--8} (\bibinfo{year}{2004}).

\bibitem{finlayson2015}
\bibinfo{author}{Finlayson, G.~D.}, \bibinfo{author}{Mackiewicz, M.} \&
  \bibinfo{author}{Hurlbert, A.}
\newblock \bibinfo{title}{Color correction using root-polynomial regression}.
\newblock \emph{\bibinfo{journal}{IEEE Trans. Image Process.}}
  \textbf{\bibinfo{volume}{24}}, \bibinfo{pages}{1460--1470}
  (\bibinfo{year}{2015}).

\bibitem{anzagira2015}
\bibinfo{author}{Anzagira, L.} \& \bibinfo{author}{Fossum, E.~R.}
\newblock \bibinfo{title}{Color filter array patterns for small-pixel image
  sensors with substantial cross talk}.
\newblock \emph{\bibinfo{journal}{J. Opt. Soc. Am. A}}
  \textbf{\bibinfo{volume}{32}}, \bibinfo{pages}{28--34}
  (\bibinfo{year}{2015}).

\bibitem{youngworth2000}
\bibinfo{author}{Youngworth, K.~S.} \& \bibinfo{author}{Brown, T.~G.}
\newblock \bibinfo{title}{Focusing of high numerical aperture
  cylindrical-vector beams}.
\newblock \emph{\bibinfo{journal}{Opt. Express}} \textbf{\bibinfo{volume}{7}},
  \bibinfo{pages}{77--87} (\bibinfo{year}{2000}).

\bibitem{luke2015}
\bibinfo{author}{Luke, K.}, \bibinfo{author}{Okawachi, Y.},
  \bibinfo{author}{Lamont, M. R.~E.}, \bibinfo{author}{Gaeta, A.~L.} \&
  \bibinfo{author}{Lipson, M.}
\newblock \bibinfo{title}{Broadband mid-infrared frequency comb generation in a
  {Si$_3$N$_4$} microresonator}.
\newblock \emph{\bibinfo{journal}{Opt. Lett.}} \textbf{\bibinfo{volume}{40}},
  \bibinfo{pages}{4823--4826} (\bibinfo{year}{2015}).

\bibitem{malitson1965}
\bibinfo{author}{Malitson, I.~H.}
\newblock \bibinfo{title}{Interspecimen comparison of the refractive index of
  fused silica}.
\newblock \emph{\bibinfo{journal}{J. Opt. Soc. Am.}}
  \textbf{\bibinfo{volume}{55}}, \bibinfo{pages}{1205--1209}
  (\bibinfo{year}{1965}).

\bibitem{cover2006}
\bibinfo{author}{Cover, T.~M.} \& \bibinfo{author}{Thomas, J.~A.}
\newblock \emph{\bibinfo{title}{Elements of Information Theory}}
  (\bibinfo{publisher}{Wiley-Interscience}, \bibinfo{address}{Hoboken, NJ},
  \bibinfo{year}{2006}), \bibinfo{edition}{2} edn.

\bibitem{lindley1956}
\bibinfo{author}{Lindley, D.~V.}
\newblock \bibinfo{title}{On a measure of the information provided by an
  experiment}.
\newblock \emph{\bibinfo{journal}{Ann. Math. Stat.}}
  \textbf{\bibinfo{volume}{27}}, \bibinfo{pages}{986--1005}
  (\bibinfo{year}{1956}).

\bibitem{vercruysse2019}
\bibinfo{author}{Vercruysse, D.}, \bibinfo{author}{Sapra, N.~V.},
  \bibinfo{author}{Su, L.}, \bibinfo{author}{Trivedi, R.} \&
  \bibinfo{author}{Vu{\v{c}}kovi{\'c}, J.}
\newblock \bibinfo{title}{Analytical level set fabrication constraints for
  inverse design}.
\newblock \emph{\bibinfo{journal}{Sci. Rep.}} \textbf{\bibinfo{volume}{9}},
  \bibinfo{pages}{8999} (\bibinfo{year}{2019}).

\bibitem{hammond2021}
\bibinfo{author}{Hammond, A.~M.}, \bibinfo{author}{Oskooi, A.},
  \bibinfo{author}{Johnson, S.~G.} \& \bibinfo{author}{Ralph, S.~E.}
\newblock \bibinfo{title}{Photonic topology optimization with
  semiconductor-foundry design-rule constraints}.
\newblock \emph{\bibinfo{journal}{Opt. Express}} \textbf{\bibinfo{volume}{29}},
  \bibinfo{pages}{23916--23938} (\bibinfo{year}{2021}).

\end{thebibliography}
\endgroup

\clearpage
\setcounter{section}{0}
\setcounter{figure}{0}
\setcounter{table}{0}
\setcounter{equation}{0}
\setcounter{page}{1}
\renewcommand{\thesection}{S\arabic{section}}
\renewcommand{\thefigure}{S\arabic{figure}}
\renewcommand{\thetable}{S\arabic{table}}
\renewcommand{\theequation}{S\arabic{equation}}
\renewcommand{\thepage}{S\arabic{page}}
\begin{center}
{\LARGE\bfseries Supplementary Information}\\[6pt]
{\large Imaging-system-aware color routers optimized for imaging information}\\[4pt]
{Hyoseok Park, Sehyeon Park, Myungjae Lee and Yeonsang Park}
\end{center}
\vspace{1em}

\section{The imaging-system-pupil input source}
\label{si:source}

\begin{figure}[tbp]
\centering
\includegraphics[width=0.62\linewidth]{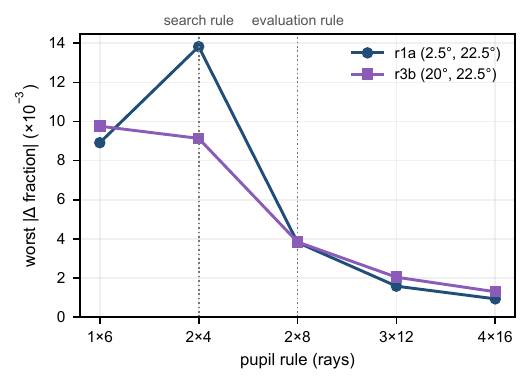}
\caption{\textbf{The adopted pupil quadrature is converged.} Worst absolute
well-fraction shift, over sites and wavelengths, against a dense $5\times20$
($100$-ray) reference as the tilted-pupil rule is refined from six to
sixty-four rays, for a near-axis (r1a) and an off-axis (r3b) mask. Marked are
the $2\times4$ eight-ray search rule and the $2\times8$ sixteen-ray evaluation
rule used for every reported value, where the shift is already
below $5\times10^{-3}$.}
\label{fig:Squadrature}
\end{figure}

\begin{figure}[tbp]
\centering
\includegraphics[width=0.86\linewidth]{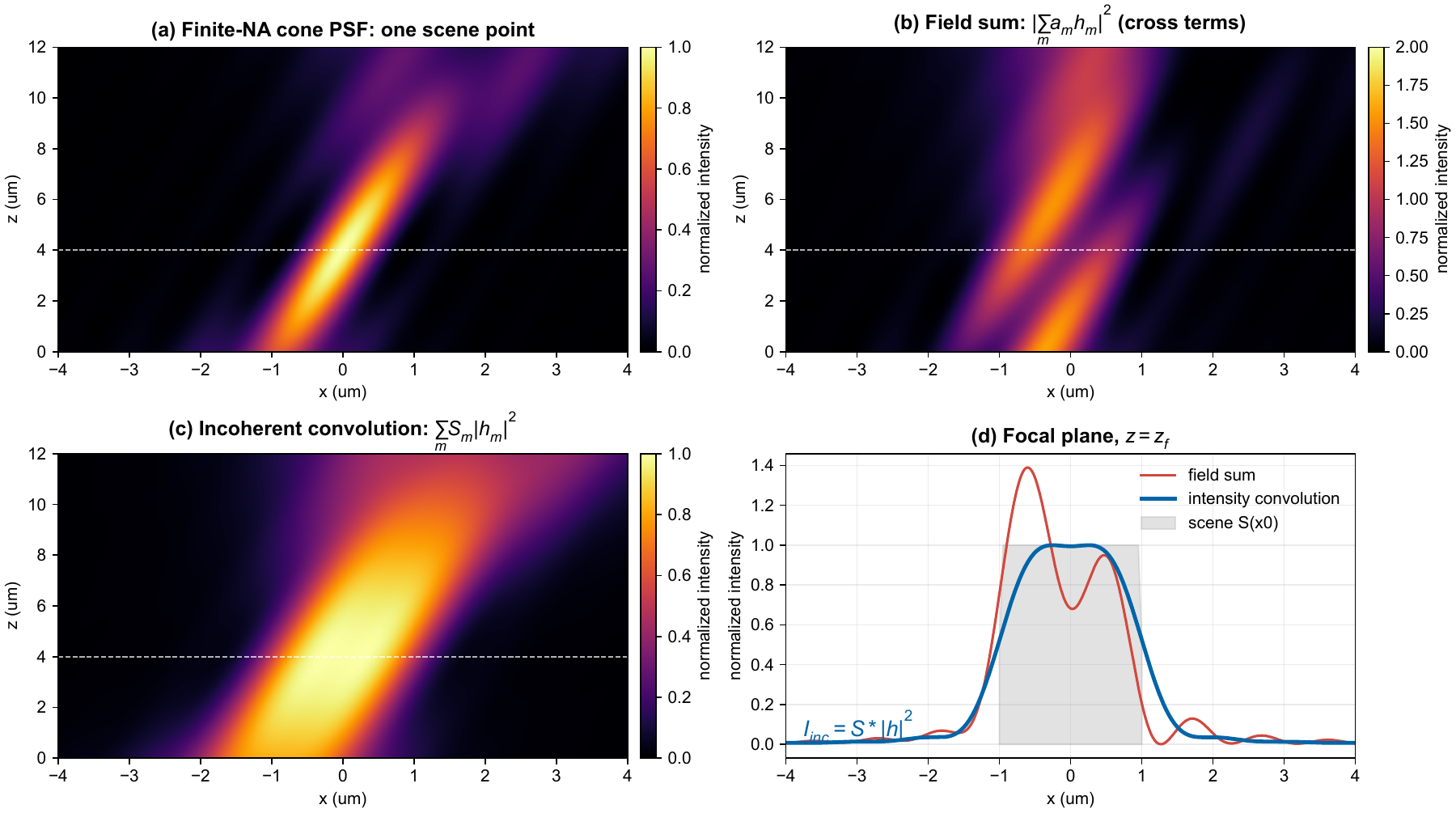}
\caption{\textbf{Coherent field sum versus incoherent intensity convolution.}
(a) The finite-NA cone point-spread function of one scene point.
(b) The pupil components summed as fields, $|\sum_m a_m h_m|^2$, retaining the
cross terms: the coherent focal model. (c) The same components summed as
intensities, $\sum_m S_m|h_m|^2$: the incoherent convolution an extended scene
delivers. (d) Focal-plane comparison of the two against the scene itself.
Propagation spans the full cell depth.}
\label{fig:Sincoh}
\end{figure}

The imaging-system-pupil illumination is represented by a precomputed geometric
library independent of wavelength, quantum efficiency and device structure.
Its reduction to the plane-wave ensemble has one physical stage and one
numerical stage. Physically, a single scene point imaged to focus position
$\mathbf x_0$ produces the coherent cone
\begin{equation}
\bm E_{\mathbf x_0}(\rperp)=\int_P\sqrt{w(\shat)}\;\bm E_{\shat}(\rperp)\,
e^{-i\kperp(\shat)\cdot\mathbf x_0}\,d^2\shat,
\label{eq:Scone}
\end{equation}
with $\bm E_{\shat}$ the rigorous coupled-wave response of the cell to pupil
direction $\shat$ and $P$ the pupil support. A detected site power is quadratic
in the field, so that one cone contributes
\begin{equation}
P_b(\mathbf x_0)=\int_P w(\shat)\!\int_{\Omega_b}\!\left|\bm E_{\shat}\right|^2
+\iint_{\shat\ne\shat'}\!\!\sqrt{w(\shat)w(\shat')}\;
e^{-i\left[\kperp(\shat)-\kperp(\shat')\right]\cdot\mathbf x_0}
\int_{\Omega_b}\!\bm E_{\shat}\!\cdot\!\bm E^{*}_{\shat'},
\label{eq:Squadratic}
\end{equation}
in which $\Omega_b$ is the area of site $b$. For a spatially incoherent scene
the focus position is uniformly distributed over many cell periods, and
averaging over $\mathbf x_0$ annihilates every off-diagonal phase, leaving
\begin{equation}
\bar P_b(\lambda)=\int_P w(\shat)\,P_b(\shat;\lambda)\,d^2\shat.
\label{eq:Sensemble}
\end{equation}
Equation~\eqref{eq:Sensemble} is an identity rather than an approximation under
the conditions it names: a scene spatially incoherent and uniform over a patch
large compared with the coherence radius $0.61\lambda/\mathrm{NA}$, which runs
from $0.92$ to $1.32\,\mu$m across the design band, imaged by a lens
shift-invariant over the cell onto an array of identical periodic cells.
Summing the pupil components as fields would retain cross terms that summing
them as intensities does not (Fig.~\ref{fig:Sincoh}); that difference is the
interference structure a coherent-cone-optimized design exploits and the scene
average deletes, so a design certified against the coherent sum does not
transfer to an imaging system.
Numerically, that
continuous pupil integral is sampled by a finite cubature,
$\bar P_b\approx\sum_{n=1}^{16}w_nP_b(\shat_n)$ for the design rule. The
number of rays is therefore a \emph{quadrature} count, not a physical count of
cones: the scene delivers infinitely many mutually incoherent cones, whose
average is a continuous intensity mixture over the pupil, and eight is how
finely that continuum is sampled for gradients, sixteen for every reported
value; a fourfold refinement of the rule leaves the routed fractions unchanged
(Fig.~\ref{fig:Squadrature}).

\paragraph{Quadrature on the tilted disk.}
A field point is specified by the chief-ray magnitude $\theta_{\mathrm{CRA}}$
and azimuth $\phi_{\mathrm{CRA}}$. The exit-pupil support is the
$\mathrm{NA}=0.30$ disk in direction-cosine space, rigidly translated by
$\sin\theta_{\mathrm{CRA}}\,\hat{\bm u}(\phi_{\mathrm{CRA}})$. A sine-condition
system would additionally foreshorten the off-axis pupil into an ellipse
compressed by $\cos\theta_{\mathrm{CRA}}$ along the tilt, $9.4\%$ at the corner;
the disk is kept circular here, so the sampled pupil area is held fixed as the
field point moves. It is sampled by
a deterministic polar quadrature of $n_r$ radial by $n_\phi$ azimuthal nodes.
Every reported value uses $\{2,8\}$, i.e.\ $16$ rays. The differentiable
search runs on the $\{2,4\}$ eight-ray reduction of the same rule, which is
cheaper by half. On the accepted on-axis mask the reduction moves the score by
$0.030$~bit and the site fractions by $0.014$, so designs are searched on it and
scored on the full rule, which itself holds the score to $0.0035$~bit and the
routed fractions to $3.5\times10^{-3}$ against the four-fold refinement
$\{4,16\}$.
Each node carries the weight $w_n=A(\theta_n,\phi_n)\Delta A_n/\cos\theta_n$ of
the main text's pupil-weight definition, in which $\theta_n$ is the polar angle measured from
the \emph{sensor normal} after the tilt shift, not from the chief ray. Because
$\Delta A_n/\cos\theta_n$ is the solid-angle element, the construction is a
pupil of uniform intensity, equal power per unit solid angle. This convention produces
an apodization asymmetry across the disk at oblique field points, with the far edge carrying more weight than the near
edge.

Every ray is solved for two orthogonal Cartesian incident polarizations and the
detected powers are averaged with equal weight, giving the unpolarized response.
Unpolarized operation is a precondition of the improper (mirror) elements of the
deployment group in Sec.~\ref{si:orbit}.

All pupil directions, the disk radius and the chief-ray angle are specified on
the air (lens) side. The flat superstrate interface refracts each direction
according to $n_{\mathrm{sup}}\sin\theta_{\mathrm{sup}}=\sin\theta_{\mathrm{air}}$,
which conserves the transverse wavevector, so the in-medium transverse
wavevector is $\kperp=k_0(s_x,s_y)$ with $(s_x,s_y)$ the air-side direction
cosines. All chief-ray angles are air-side.

The weights are normalized so that $\sum_n w_n=\mathcal R(\theta_{\mathrm{CRA}})$,
a declared relative-illumination model. The present implementation uses the
conventional $\cos^4$ falloff for an ideal lens; modules with shifted
microlenses commonly fall off more slowly \cite{catrysse2000qe,vaillant2008,huo2010},
and a CR, which removes the microlens, has no pupil-matching element at all.
Both the CR and the CFA baseline carry the identical factor, so a comparison
at a fixed field point is insensitive to the choice.

Cone diffraction, aperture convolution, neighboring-cell point-spread spill and
the chief-ray tilt are all contained in the single weight set
$\{(\theta_n,\phi_n),w_n\}$. The spill is contained only because the
construction is flat-field and infinite-periodic: every cell is identical and
illuminated identically, so spill into a cell equals spill out of it and
the two cancel.

\section{The electromagnetic forward model}
\label{si:forward}

Each pupil ray at each wavelength is one independent rigorous coupled-wave
analysis solve \cite{moharam1981,moharam1995a,moharam1995b} with the
Fourier-factorization and numerical-stability refinements of
Refs.~\cite{li1996a,li1996b}, evaluated in the
differentiable GPU implementation TORCWA \cite{kim2023}. The cell is a
$2\,\mu$m four-site supercell carrying one air-clad SiN design layer of
common height $600$~nm on an oxide substrate, with measured dispersion for SiN \cite{luke2015} and
SiO$_2$ \cite{malitson1965}. A back-illuminated stack \cite{wakabayashi2010,fossum2014}
would ordinarily planarize this layer in oxide, which reduces the index contrast
relative to the air-clad structure simulated here and would correspondingly weaken
the achievable routing.

Per-order transmitted power is accumulated as
\begin{equation}
T(\lambda,\theta,\phi)=\!\!\sum_{mn\ \text{prop.}}\!\!
\frac{\mathrm{Re}\,k_{z,t}^{(mn)}}{k_{z,i}}
\bigl(|t^{s}_{mn}|^{2}+|t^{p}_{mn}|^{2}\bigr).
\end{equation}
The model evaluates the transmitted side only; the four well fractions are rescaled to this $T$.
The per-order transmitted-power expression is evaluated in the $s$--$p$
eigenpolarization basis of each transmitted order.

Reported calculations use Fourier orders $(8,8)$ on a $128\times128$ density grid
with a $24\times24$ detector-plane midpoint grid. A fractional frequency nudge of
$10^{-4}$ moves the solve off the Rayleigh anomaly, where an order becomes
grazing; when a solve there returns an unphysical transmittance the nudge is
retried at $0$, $5\times10^{-4}$, $10^{-3}$ and $2\times10^{-3}$. Selected hard masks are re-solved on a
convergence ladder through order $(12,12)$ with a $64$-sample
detector grid, at both the design field point and a stressed oblique field
point, across all optimizer wavelengths and the full pupil; the shifts this
produces are quantified below.

Detector-plane routed powers follow from $P_b=\int_{\Omega_b}|\bm E(\rperp)|^2
d^2\rperp$, integrated over the full quadrant at a photodiode plane
$1.25\,\mu$m below the exit interface of the design layer, inside the SiO$_2$
output medium. The four values are quadrant-integrated detector-plane intensity
fractions rescaled so that their sum equals the power-normalized
transmission above. The proxy assumes a full fill factor with no deep-trench
isolation, partitions detector-plane intensity rather than resolving absorption
or charge collection in silicon, and therefore constitutes an optical upper
estimate to the extent that a real photodiode's active area is smaller than its
quadrant. The word \emph{electron} throughout this work denotes the common
quantum-efficiency and exposure conversion applied to that optical proxy, not a
device-level photodiode model.

Responses are produced in physical well order (Fig.~\ref{fig:well_layout}) and
then permuted into the fixed phase order consumed by the information model. By Sec.~\ref{si:invariance} the
objective is invariant to that ordering. Gradients are taken by reverse-mode
differentiation through the solver.

\paragraph{Fourier truncation.} The production rule keeps $(8,8)$ orders.
Against a $(16,16)$ reference the site fractions move by at most $0.029$ at
normal incidence and $0.015$ at twenty degrees, and $(12,12)$ is already
converged to $0.004$. In the reported quantity the shift is far smaller: at the
adopted settings the on-axis mask scores $\Iimg=1.6321$ at $(8,8)$ against
$1.6355$ at $(12,12)$, a difference of $0.003$~bit against an on-axis deficit of $0.22$~bit.

\section{Closed form and properties of the imaging information}
\label{si:closedform}

Let $\mathbf X\in\mathbb R^{9}$ be the measured spectral latent, $\ZXYZ\in
\mathbb R^{3}$ the full-CIE target, and let the scene prior specify the
joint second moments $\Cov{X}$, $\Cov{XZ}$ and
$\Cov{Z}$ of the zero-mean pair $(\mathbf X,\ZXYZ)$. The four
raw measurements are $\mathbf Y=\Amat\mathbf X+\mathbf n$ with
$\mathbf n\sim\mathcal N(\bm 0,\Sn)$ independent of $(\mathbf X,\ZXYZ)$. Then
$(\mathbf Y,\ZXYZ)$ is jointly Gaussian with
\begin{equation}
  \Cov{Y}=\Amat\Cov{X}\Amat^{\mathsf T}+\Sn,
  \qquad
  \Cov{YZ}=\Amat\Cov{XZ},
\end{equation}
so the conditional law of $\ZXYZ$ given $\mathbf Y$ is Gaussian with covariance
\begin{equation}
  \Cov{Z|Y}
  =\Cov{Z}
  -\Cov{ZX}\Amat^{\mathsf T}\Cov{Y}^{-1}
   \Amat\Cov{XZ} .
  \label{eq:si_posterior}
\end{equation}
For jointly Gaussian variables the mutual information is the difference of
differential entropies, $I(\mathbf Y;\ZXYZ)=h(\ZXYZ)-h(\ZXYZ\mid\mathbf Y)$
\cite{cover2006}, and a Gaussian in $d$ dimensions with covariance $\mathbf{C}$
has $h=\tfrac12\log_2\!\left[(2\pi e)^{d}\det\mathbf{C}\right]$. The dimensional
factors cancel, leaving
\begin{equation}
  I(\mathbf Y;\ZXYZ)=\frac{1}{2}\log_2
  \frac{\det\Cov{Z}}
       {\det\Cov{Z|Y}} ,
  \label{eq:si_iimg}
\end{equation}
for the four-site cell, so that $\Iimg=\tfrac14 I(\mathbf Y;\ZXYZ)$ is the
per-raw-pixel form quoted in the main text. Because $\Cov{Z}$ does not
depend on the structure, maximizing $\Iimg$ is equivalent to minimizing
$\det\Cov{Z|Y}$: the design minimizes the volume of the
posterior color uncertainty ellipsoid. This is the D-optimality criterion of
optimal experimental design \cite{lindley1956}.

Equation~\eqref{eq:si_iimg} is evaluated through Cholesky factorizations, with
the posterior assembled from a whitened latent information matrix and the log
determinants accumulated from triangular factors. No explicit matrix inverse is
formed, which matters because $\Cov{Y}$ becomes poorly conditioned at high
exposure, where the shot-noise floor is small relative to the signal covariance.

The expression is exact for the stated linear--Gaussian model, two of whose
ingredients approximate the physical imaging system. The Poisson photon statistics are
represented by their Gaussian equivalent, $\Sn=\operatorname{diag}(\bar{\mathbf y})+\sigma_{\mathrm r}^2
\mathbf I$, in which $\bar{\mathbf y}$ is the expected (exposure-mean) count, so
$\Sn$ is fixed and realization-independent and $\mathbf n$ remains independent of
$(\mathbf X,\ZXYZ)$; this expected-count noise is accurate at the electron counts
considered but is not the exact count distribution, and a per-realization,
signal-dependent shot covariance would not leave $(\mathbf Y,\ZXYZ)$ jointly
Gaussian; and the scene prior is a second-moment description of
real reflectance and illuminant data rather than a claim that natural spectra
are Gaussian.

\paragraph{The irreducible target term.}
\label{si:nuisance}
The measurement latent $\mathbf X$ is sampled inside the band the optical model
covers, while the target $\ZXYZ$ is formed on the full color-matching support
\cite{cie0152018}, so the two are not related by a deterministic map. Define
the residual target covariance
\begin{equation}
  \Cov{Z|X}
  =\Cov{Z}
  -\Cov{ZX}\Cov{X}^{-1}\Cov{XZ} ,
  \label{eq:si_nuisance}
\end{equation}
the part of the color target that \emph{no} measurement of $\mathbf X$ can
determine. Substituting an assumed deterministic relation $\ZXYZ=\mathbf T
\mathbf X$ would set Eq.~\eqref{eq:si_nuisance} to zero, and with it the
irreducible floor on the posterior. Retaining Eq.~\eqref{eq:si_nuisance} bounds
$\Iimg$ from above by $\tfrac18\log_2[\det\Cov{Z}/
\det\Cov{Z|X}]$, the information a noiseless measurement of the
entire latent would supply. That ceiling is a property of the band and the scene
prior, not of any structure, and a design's information should be read against
it. The joint moments are constructed from the measured reflectance set of
Ref.~\cite{babelcolor2012}.

\paragraph{Invariance properties.}
\label{si:invariance}
Two invariances of Eq.~\eqref{eq:si_iimg} follow from the same determinant
identity. First, color-basis invariance. Let $\ZXYZ'=\mathbf M\ZXYZ$ for any
invertible $\mathbf M\in\mathbb R^{3\times3}$,
for instance the transform to linear sRGB. Then
$\Cov{Z}'=\mathbf M\Cov{Z}\mathbf M^{\mathsf T}$
and, from Eq.~\eqref{eq:si_posterior},
$\Cov{Z|Y}'=\mathbf M\Cov{Z|Y}
\mathbf M^{\mathsf T}$. The factors $(\det\mathbf M)^2$ cancel in the ratio, so
$\Iimg$ is unchanged.

Second, recombination invariance. Let $\mathbf Y'=\mathbf P\mathbf Y$ for any
invertible $\mathbf P\in
\mathbb R^{4\times4}$. Then $\Amat'=\mathbf P\Amat$ and $\Sn'=\mathbf P\Sn
\mathbf P^{\mathsf T}$, and the correction term in Eq.~\eqref{eq:si_posterior}
becomes
\begin{equation}
  \Cov{ZX}\Amat^{\mathsf T}\mathbf P^{\mathsf T}
  \left(\mathbf P\Cov{Y}\mathbf P^{\mathsf T}\right)^{-1}
  \mathbf P\Amat\Cov{XZ},
\end{equation}
in which $\mathbf P^{\mathsf T}(\mathbf P\Cov{Y}\mathbf P^{\mathsf T})^{-1}
\mathbf P=\Cov{Y}^{-1}$ exactly. The posterior, and hence $\Iimg$, is
therefore invariant under \emph{any} invertible linear recombination of the four
raw channels, of which a permutation of the wells is the special case
$\mathbf P$ a permutation matrix. The four detector sites therefore need no
color identity, and labeling them $q_1,\dots,q_4$ costs nothing. The lattice
isometries used for field deployment induce exactly such a permutation, so the
eight copies of an optimized mask are scored identically.

\section{Deployment orbit arithmetic}
\label{si:orbit}

Acting on the tilt azimuth, the dihedral group of the square sends
$\phi\mapsto\pm\phi+90^\circ k$. The orbit of a design azimuth $\phi_c$ is
therefore
\begin{equation}
  \{\pm\phi_c,\;90^\circ\pm\phi_c,\;180^\circ\pm\phi_c,\;270^\circ\pm\phi_c\},
\end{equation}
which contains eight distinct azimuths for a generic $\phi_c$ (four mirror
pairs), collapses to four spaced by $90^\circ$ when $\phi_c$ lies on a symmetry
axis, and is an equally spaced $45^\circ$ comb only in the special case
$\phi_c=22.5^\circ\ (\mathrm{mod}\ 45^\circ)$. Wedge centers must consequently
be chosen so that their orbits tile each ring exactly; a uniform $45^\circ$ comb does not generally reproduce the $D_4$ orbit and leaves azimuths served off-axis by more than the intended
sector half-width.

An azimuth mismatch $\Delta\phi$ displaces the pupil-disk center along the
chord
\begin{equation}
  |\Delta\bm s| = 2\sin\theta_{\mathrm{CRA}}\,\sin(\Delta\phi/2)
  \;=\; \sin\theta_{\mathrm{CRA}}\,\Delta\phi + O(\Delta\phi^3),
  \label{eq:si_chord}
\end{equation}
so the same angular sector displaces the source further, in the units the optics
cares about, as the chief-ray angle grows. Evaluated at each ring-center
chief-ray angle as a fraction of the $\mathrm{NA}=0.30$ pupil radius, the adopted
half-widths hold the center-to-edge displacement below $15\%$ in every ring,
against $44.5\%$ for a uniform $45^\circ$ sector at ring~3.

The electromagnetic half of the deployment claim rests on the
invariance of Maxwell's equations under the joint map
(mask $\to$ transformed mask, azimuth $\to$ mapped azimuth), together with the
requirement that the illumination be unpolarized and the media isotropic,
non-chiral and non-magnetic; it is not separately re-simulated. The
reconstruction half is analytic: by Sec.~\ref{si:invariance} the change in
$\Iimg$ under any of the eight well relabelings is exactly zero.

\begin{figure}[tbp]
\centering
\includegraphics[width=\linewidth]{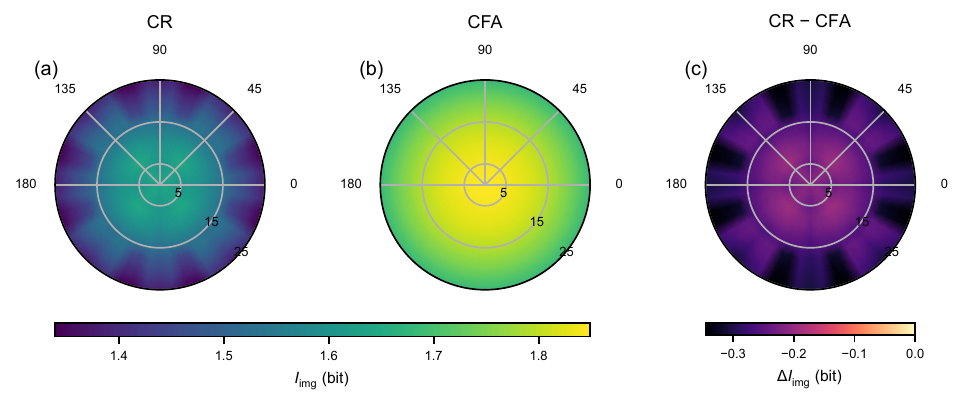}
\caption{\textbf{Imaging information across the field.} Polar maps over the
domain $(\theta_{\mathrm{CRA}},\phi)\in[0,25^\circ]\times[0,360^\circ)$, radius
the chief-ray angle and angle the tilt azimuth, evaluated from the deployed
six-mask set. \textbf{(a)}~The CR, each field point scored with the design its
deployment rule assigns; the eight-fold pattern is the azimuth dependence of the
routing, absent for the CFA. \textbf{(b)}~The CFA at the same field points,
flat in azimuth and falling with chief-ray angle through the relative
illumination. \textbf{(c)}~Their difference is negative everywhere: the CR
trails by $0.22$~bit on axis, by as little as $0.19$~bit near $7.5^\circ$ on the
diagonal, and by $0.35$~bit at the corner. The wedge below
$45^\circ$ is computed and carried to the circle by the deployment group; panels
(a) and (b) share one scale.}
\label{fig:Sfieldmap}
\end{figure}

\section{The reconstructed literature designs}
\label{si:litdesigns}

\begin{figure}[p]
\centering
\includegraphics[width=\linewidth]{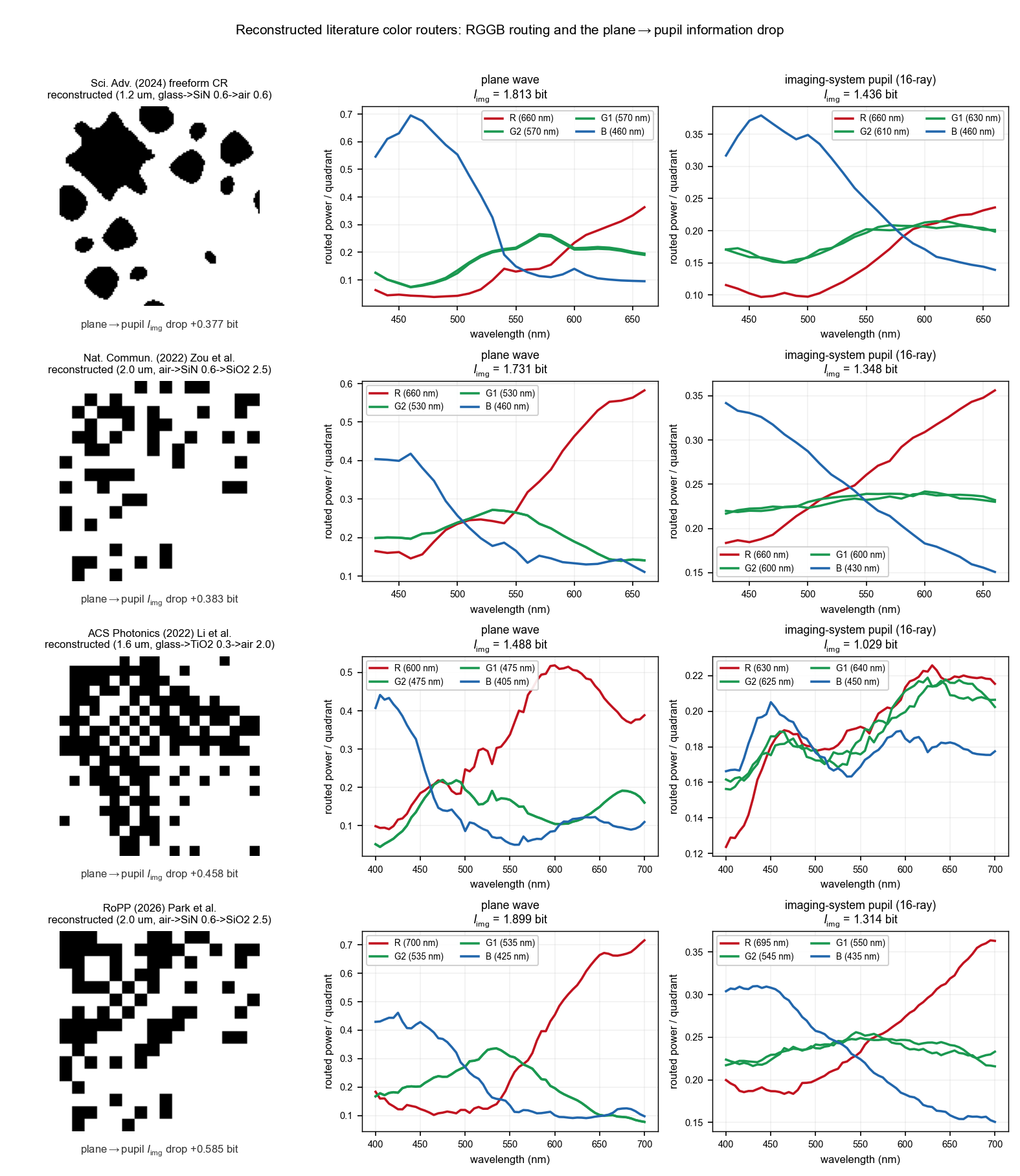}
\caption{\textbf{The four reconstructed literature color routers.} Rows, top to
bottom: the freeform design~\cite{kim2024router}, the Bayer
router of Zou et~al.~\cite{zou2022}, the ${\rm TiO_2}$ Bayer splitter of
Li et~al.~\cite{li2022bayer}, and the deep-learning-designed
router of Park et~al.~\cite{park2026ropp}. Left: the reconstructed mask, with the layer
stack used for each. Center: per-well routed spectra under a single normal
plane wave, the source each publication optimizes and reports against, with
the hard-mask imaging information. Right: the same design under the
sixteen-ray imaging-system-pupil ensemble used throughout this work.}
\label{fig:Sexternal}
\end{figure}

Three of the four literature points -- the freeform color router~\cite{kim2024router},
the Bayer-pattern router of Zou et~al.~\cite{zou2022}, and the ${\rm TiO_2}$ Bayer
splitter of Li et~al.~\cite{li2022bayer} -- are not simulated from a
published permittivity file, which none of those papers releases, but reconstructed from the
design image in the source publication: each mask was rasterized from the published
figure at the reported cell pitch and layer stack, then validated by reproducing that
paper's own plane-wave RGGB routing. The fourth, the deep-learning-designed router of
Park et~al.~\cite{park2026ropp}, is our own work, so we use its plane-wave routing from
that work's full-wave (FDTD) data directly. All four are then scored under the pupil
rule used throughout this work (Fig.~\ref{fig:Sexternal}).
Re-scored under the imaging-system-pupil ensemble, the four designs lose
$0.38$, $0.38$, $0.46$, and $0.59$~bit of imaging information respectively. Figure~\ref{fig:rescored}
shows the same non-transfer across the full set of $109$ plane-wave-optimized
routers.

\begin{figure}[tbp]\centering
\includegraphics[width=\linewidth]{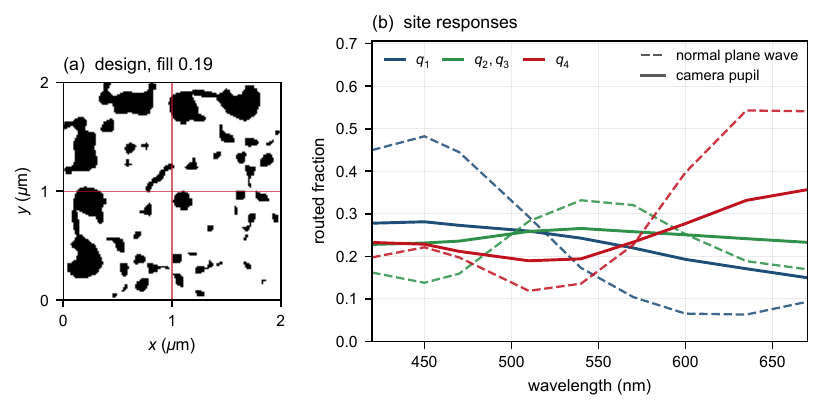}
\caption{\textbf{The plane-wave routing does not survive the pupil ensemble.}
\textbf{(a)}~The design itself, viewed from above: black is SiN and white the air
within the layer, and the red lines mark the four detector sites beneath.
\textbf{(b)}~Its four site responses, evaluated at its own geometry under the
normal plane wave it was optimized against (dashed) and under the on-axis
illumination ensemble (solid). Under the plane wave the cell sorts into a blue, two green
and a red allocation; the two diagonal wells are mapped onto each other by the
$y=x$ mirror and carry the same spectrum to $6\times10^{-3}$, so they are drawn
once. Under the ensemble every well relaxes toward the flat quarter, and
$\Iimg$ falls from $1.95$ to $1.32$\,bit. The design shown is the
highest-scoring of the nine mirror-symmetric designs among the $109$
plane-wave-optimized routers. Routed fractions are optical power
fractions per site, before quantum-efficiency weighting.}
\label{fig:ourplanewave}
\end{figure}

\begin{figure}[tbp]
\centering
\includegraphics[width=\linewidth]{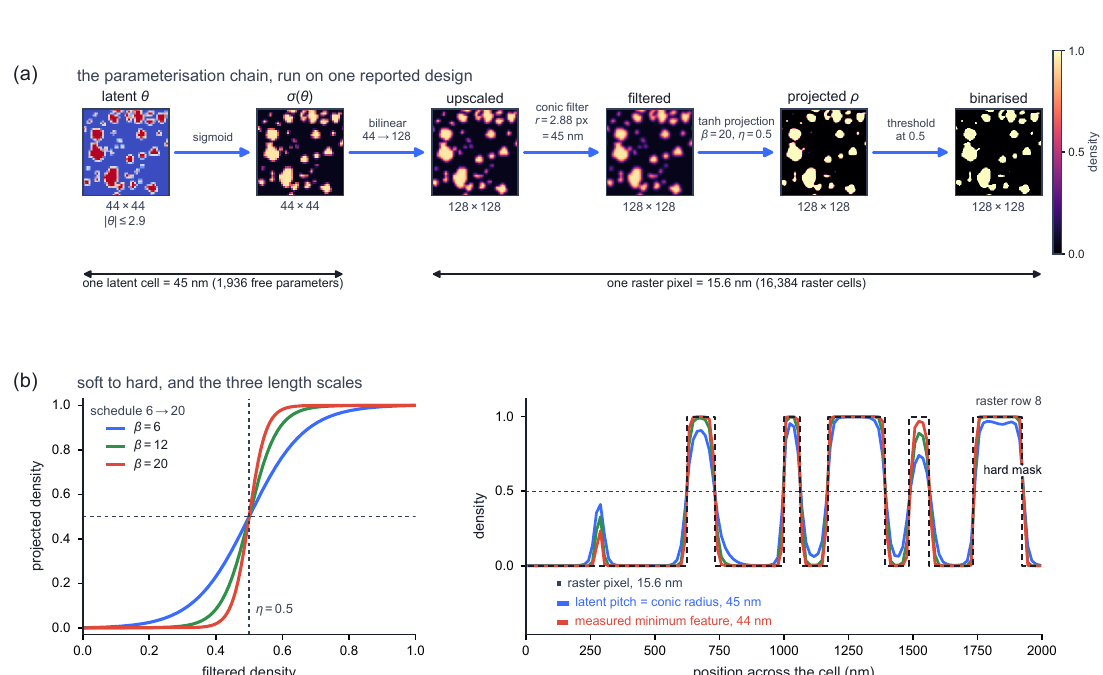}
\caption{\textbf{The parameterization and its length scales.}
\textbf{(a)}~The chain on one reported design: a $44\times44$ latent field
through a sigmoid, bilinear upscaling to the $128\times128$ raster, a conic
filter of radius $45$\,nm, a tanh projection at $\beta=20$, $\eta=0.5$, and a
threshold. \textbf{(b)}~The soft-to-hard projection over the $\beta$ schedule,
and one raster row showing the three length scales: the raster pixel
($15.6$\,nm), the latent pitch equal to the conic radius ($45$\,nm), and the
measured minimum feature ($44$\,nm).}
\label{fig:parameterization}
\end{figure}

\begin{figure}[tbp]\centering
\includegraphics[width=0.38\linewidth]{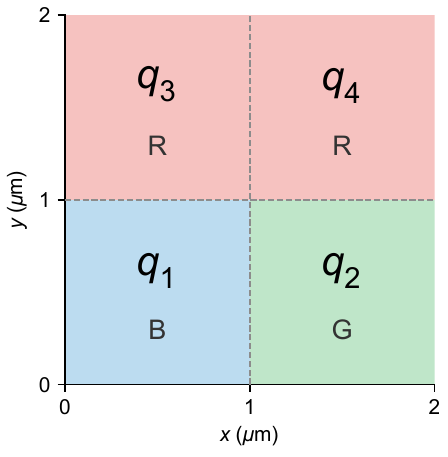}
\caption{\textbf{Detector-site layout of the supercell.}
The four sites $q_1$--$q_4$ of the $2\,\mu$m cell and the spectral band each site
predominantly collects under the on-axis pupil ensemble.}
\label{fig:well_layout}
\end{figure}

\begin{figure}[tbp]\centering
\includegraphics[width=\linewidth]{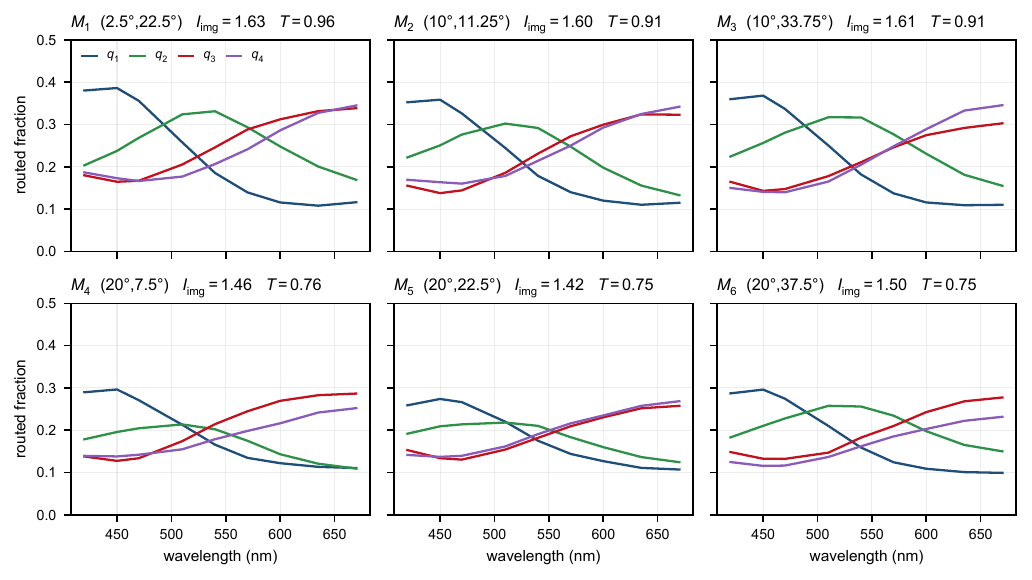}
\caption{Spectral responses of the six designed CRs at their design
chief-ray angles and azimuths. Each panel shows the response of the four
detector sites $q_1$--$q_4$; the title reports the $\Iimg$ recomputed from that
same response, reproducing the per-slot column of the main text's six-mask
deployment figure, together with the band-averaged summed throughput $T$. The responses are
broadband and mutually overlapping, and $T$ falls from $0.96$ on axis to $0.75$
at twenty degrees.
Site responses are optical power fractions before quantum-efficiency weighting.}
\label{fig:champ_spectra}
\end{figure}

\begin{figure}[tbp]\centering
\setlength{\fboxsep}{6pt}%
\fbox{\includegraphics[width=0.60\linewidth]{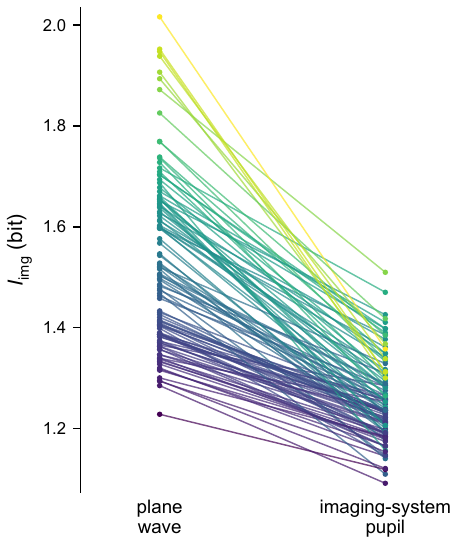}}
\caption{\textbf{The plane-wave ranking does not survive the change of source.}
The $109$ plane-wave-optimized single-layer color routers of Fig.~1(d,e), each re-scored under the normal plane wave (left) and under the imaging-system-pupil ensemble (right) under bright light (the exposure-ladder-averaged $\Iimg$) and drawn as one line per design, coloured by plane-wave score. The lines cross, so a design's plane-wave rank does not predict its imaging-system-pupil rank.}
\label{fig:rescored}
\end{figure}

\section{The design method}
\label{si:method}

The design objective is the equally weighted mean of $\Iimg$ over the five
reference-CFA-green exposure levels $\{20,40,80,160,320\}$~e$^-$ of the main
text, averaged after the logarithm rather than by averaging posteriors and taking
one logarithm. The two ends of the ladder are governed by different physics: at
the photon-starved end the read term is a substantial fraction of the total
variance and the design benefit is collection, whereas at the bright end the
shot term dominates and the benefit is spectral conditioning of the demix. A
single exposure would let the optimizer specialize to one regime; the equally
weighted mean over the ladder makes a design that is excellent at one level
and poor at another no better than one that is uniformly adequate.
The exposure scale itself is fixed once, from the area-weighted CFA green
response of the reference imaging system integrated on the measurement quadrature, and
all CR and CFA responses retain a common absolute electron-domain scale without per-device normalization.

The optimized variable is a $44\times44$ latent field
(Fig.~\ref{fig:parameterization}), bilinearly upscaled to the
$128\times128$ permittivity grid, passed through a conic filter of radius
$45$\,nm and a tanh projection in the robust erosion/dilation formulation of the main text. The free parameters number
$1{,}936$ against the $16{,}384$ of the raster, and the minimum feature follows
from the parameterization rather than from the filter alone.

The forward pass is straight-through: the thresholded binary mask is what the
solver sees and what is scored, while the gradient is taken through the relaxed
density. The two differ: at the lowest projection sharpness the relaxed
density scores $0.686$~bit below the mask that is scored, so the projection
sharpness follows a schedule from $6$ to $20$ over the run rather than being
held at its lowest value. The latent is updated by Adam at a learning rate of
$0.03$ for $100$ to $150$ steps.

Adam proposes each step, and a hard-mask evaluation admits it only if the
hard mask it produces scores above the accepted one under the search
quadrature; otherwise the latent and the optimizer moments revert to the
accepted state and the learning rate is halved, recovering on the next
acceptance.
Acceptance is tested every fourth step, so the search between tests is
unconstrained. Runs are warm-started from an existing mask, whose latent is
recovered by fitting the coarse field to the target pattern before the run
begins.

Search uses the eight-ray reduction and every reported value the sixteen-ray
rule; the eight-ray figures quoted in this section are search proxies.

Each pupil direction is an independent RCWA solve at two polarizations, so per
step the eight-ray search and the sixteen-ray evaluation cost roughly eight and
sixteen times a single normal-incidence solve. The reverse-mode gradient also
retains the intermediate fields of every direction and wavelength, and for the
full ensemble this approaches the memory of a single GPU, so the pupil
optimization trades memory for compute through gradient checkpointing and its
recomputation, which raises the effective per-step cost further above the
nominal ray count. The plane-wave objective carries neither penalty and is
cheap to sample broadly, so the plane-wave designs in Fig.~1(d,e) of the main
text ($109$ in total) span a range of layer heights and detector gaps at
several seeds, whereas the imaging-system-pupil objective is far more costly per design
and was optimized at only a small number of configurations.

The CFA reference is a spectral transmittance calculation that does not
propagate through the CR-to-detector gap; that asymmetry is physical, since a dye CFA sits in contact with the pixel.

\section{Fabrication tolerance}
\label{si:integrity}

Perturbing the reported mask by one raster pixel of critical dimension
($\pm16$\,nm erosion or dilation, about a third of the $45$\,nm design
feature) and by $\pm5\%$ of height moves the
throughput by at most $0.008$ but costs up to $0.128$\,bit of imaging
information, three fifths of the on-axis deficit against the CFA, dropping the effective
purity from $0.17$ to $0.12$ while the collected power is unchanged. The
height change alone moves the routed fractions about six times less than the
one-pixel bias (Sec.~\ref{si:experiment}), so the process requirement falls
on critical dimension, not on layer thickness.

Free-form optimization on a fine grid readily produces sub-fabrication features
\cite{vercruysse2019,hammond2021,zhou2015}. The $44\times44$ latent field with
robust erosion/dilation bounds the design's spatial-frequency content at one
latent cell, $45$\,nm, but that bound applies to the parameterization and not to
printed geometry: a projection threshold can place a level crossing arbitrarily
close to a latent node, leaving isolated structures a single raster pixel across.
The masks reported here have every such island of solid and hole of void removed.
Deleting them costs $0.001$ to $0.003$\,bit, so they are artifacts of the
threshold rather than features the design uses; deleting more is not free, with
everything up to $31$\,nm costing $0.022$ to $0.049$\,bit and up to $35$\,nm
costing $0.041$ to $0.076$\,bit.

After that removal the narrowest cross-section is $44$\,nm in four of the six
designs and $31$\,nm in the other two, where a neck inside a larger feature
survives an area threshold. Against the $600$\,nm layer those are etch aspect
ratios of $14{:}1$ and $19{:}1$.

\section{Experimental characterization}
\label{si:experiment}

\subsection{Sample and measurement}

\begin{figure}[tbp]\centering
\includegraphics[width=\linewidth]{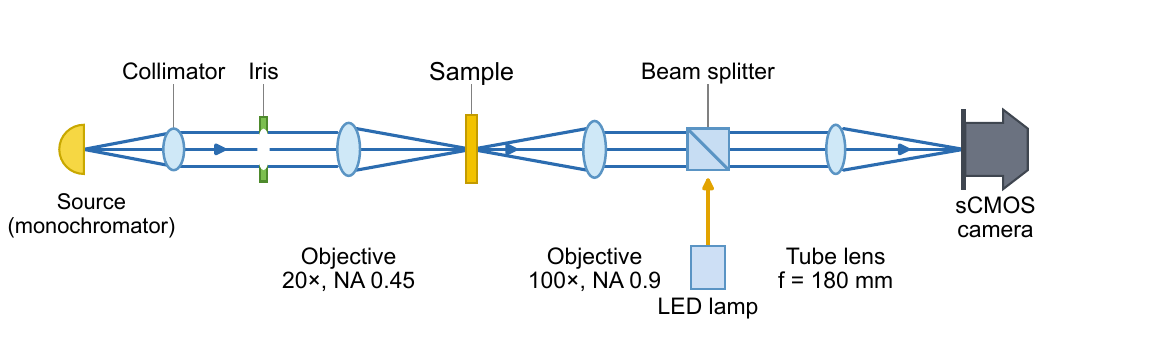}
\caption{\textbf{Measurement setup.} A collimated broadband source
(xenon lamp and monochromator) passes through an iris aperture and a
$20\times$, NA~$0.45$ condenser objective, which together set the NA~$0.3$
imaging-system pupil at the sample. The transmitted field is collected by a
$100\times$, NA~$0.9$ objective and relayed by an $f=180$\,mm tube lens through
the beam splitter onto the sCMOS sensor. For focusing, the LED lamp illuminates
the sample through the beam splitter and the light reflected from the sample
returns through the beam splitter to the camera, bringing the sample into focus
on the sensor.}
\label{fig:exp_setup}
\end{figure}

\begin{figure}[tbp]\centering
\includegraphics[width=0.86\linewidth]{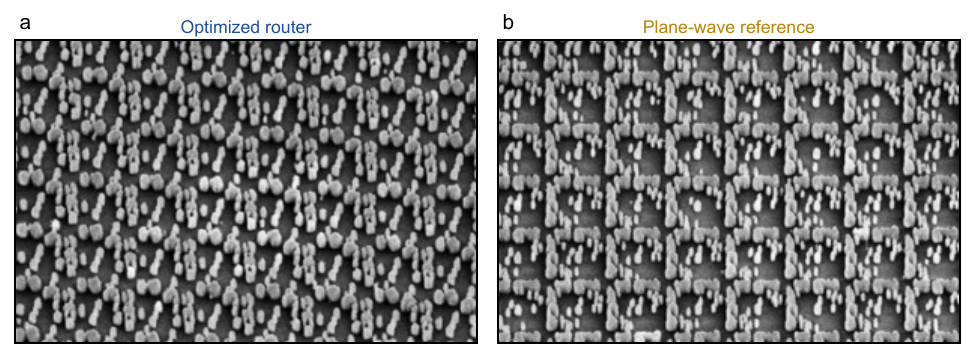}
\caption{\textbf{Tilted-view electron micrographs of the fabricated devices.}
\textbf{(a)}~Optimized router and \textbf{(b)}~plane-wave reference, imaged at a
tilt to reveal the etched sidewall. The nitride pillars stand with a clean,
near-vertical profile across the array.}
\label{fig:exp_tilt}
\end{figure}

A $600$-nm-thick silicon nitride film was deposited on a $700$-$\mu$m-thick
fused-silica substrate by plasma-enhanced chemical vapor deposition (PECVD). The
pattern was defined by electron-beam lithography (JEOL JBX-8100FS) using ZEP520A
resist coated on a conductive polymer layer and subsequently transferred to an
$80$-nm-thick evaporated chromium hard mask (SRN200I e-gun evaporator). The
nitride layer was then etched by inductively coupled plasma (ICP) etching
(Gigalane NeoS-MAXIS200H) using a CHF$_3$/O$_2$ chemistry. The fabricated samples
were illuminated from the nitride side, with light propagating toward the
fused-silica substrate. The etch depth measured $585$ to $600$\,nm against the
$600$\,nm design (Fig.~\ref{fig:exp_tilt}). A xenon lamp
feeds a monochromator that delivers narrowband light in $25$\,nm steps from
$400$ to $700$\,nm, with the source optics set to form the on-axis
($\theta_{\mathrm{CRA}}=0$) NA~$0.3$ imaging-system pupil of Sec.~S1 through an iris
aperture and a $20\times$ condenser objective
(Fig.~\ref{fig:exp_setup}). The light transmitted through the device is collected by a
$100\times$, NA~$0.9$ objective and a tube lens and recorded on a scientific CMOS
sensor (PCO edge 4.2), which samples the image at $63.6$\,nm per pixel referred
to the sample. The objective images the transmitted field a short distance beyond
the patterned layer. The focus is set once per device, coarsely from the
brightfield LED reflection of the sample and then refined against the simulated
intensity map at the design plane, and held across the band.

\FloatBarrier
\subsection{Lattice averaging}

Each recorded frame is a circular illumination spot on a dark background. A
periodic object places all of its signal on the reciprocal lattice, so we
recover one averaged supercell rather than compare pixel by pixel. We estimate
the illumination envelope by Gaussian smoothing over several lattice periods and
divide it out, locate the two shortest reciprocal-lattice vectors and reduce
them to an orthogonal basis, and project the illuminated pixels onto each
lattice harmonic out to the incoherent diffraction cutoff $2\,\mathrm{NA}/\lambda$.
The projection averages several hundred supercells within the spot, about
$500$ for the measured spot diameter. A single basis measured from the
highest-contrast frame serves the whole stack, with doubled-cell fits rejected by
the median period. Dividing out the illumination envelope removes the absolute
scale, so each measured spectrum is a per-wavelength routing fraction and the
CR--CFA crossover of main-text Fig.~3a combines those fractions with the model's
throughput.

\subsection{Alignment and uncertainty}

The mounted sample fixes one orientation and one cell origin, so we fit both once
against the as-designed structure over the eight square-lattice symmetry
operations and the full set of cyclic shifts; each frame then receives its own
translation, read from the lattice phase of the frame against the simulated
cell. The camera background is
removed with a beam-off frame, and the scattered-light pedestal, estimated from
the intensity falloff outside the illuminated spot, is subtracted before the
well powers are integrated. The effective collection aperture is fitted per
device against the simulation. The band drawn with the simulated spectra covers
the etch variants and the height variants of Sec.~\ref{si:integrity}, the
focus planes within $\pm0.5\,\mu$m of the design plane, and the
$0.80$--$0.90$ numerical-aperture range the collection aperture is fitted
over. Each of the four sources enters at its worst case and the four
multiply. We estimate the measurement uncertainty by dividing the illuminated
spot into six angular sectors and averaging each independently. The error bars
in the measured routing spectra add to that spread a one-pixel error in the
alignment, and the scattered-light correction, bracketed between removing none of
the pedestal and removing it down to the cell floor. The pedestal dominates the
three. The mean
image correlation of the averaged supercell with the as-designed simulation is
$0.66$ for the optimized router and $0.76$ for the plane-wave reference
(Fig.~\ref{fig:exp_intensity}).

\subsection{Measured supercell}

\begin{figure}[tbp]\centering
\includegraphics[width=\linewidth]{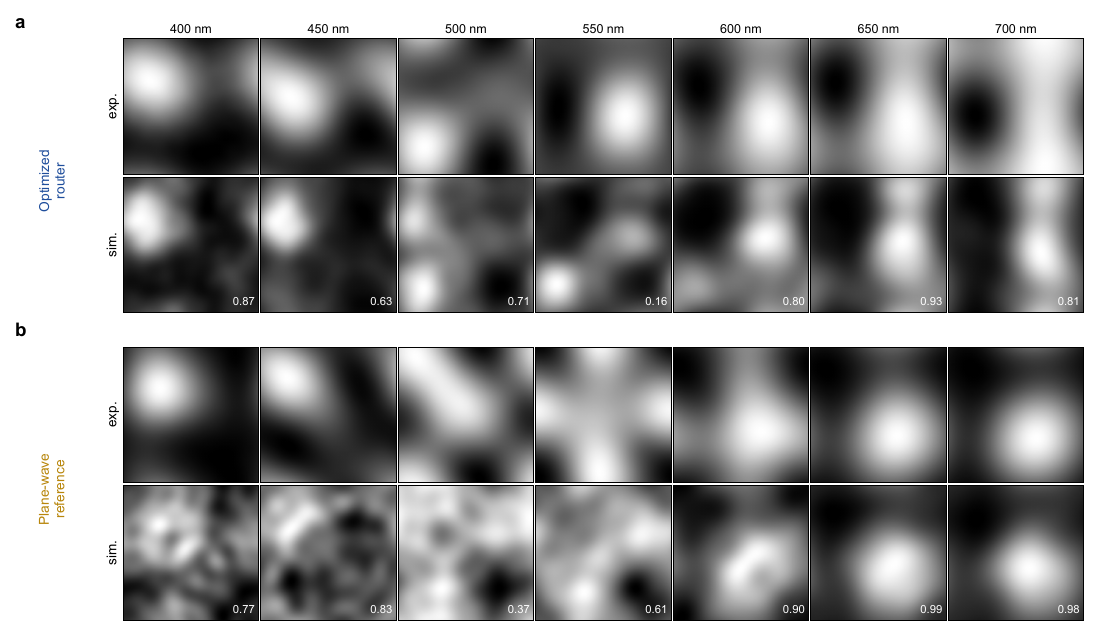}
\caption{\textbf{Measured averaged supercell against the as-designed simulation.}
\textbf{(a,b)}~Measured (exp.) and simulated (sim.) averaged supercell of the
optimized router and the plane-wave reference at seven of the thirteen measured
wavelengths, each simulated cell placed in the orientation and origin fitted once
from the as-designed structure, with the per-wavelength image correlation inset.}
\label{fig:exp_intensity}
\end{figure}

The averaged supercell of each device tracks the as-designed simulation across
the band once the simulation is placed in the single fitted pose
(Fig.~\ref{fig:exp_intensity}). The correlation is measured against the
as-designed structure rather than against whichever etch variant fits best, so
the figure reports a test rather than a fit; it is not an absolute image-quality
metric but a relative agreement that carries the combined fabrication deviation,
forward-model approximation and measurement noise.

\FloatBarrier
\subsection{Fabricated pattern}

\begin{figure}[htbp]\centering
\includegraphics[width=0.86\linewidth]{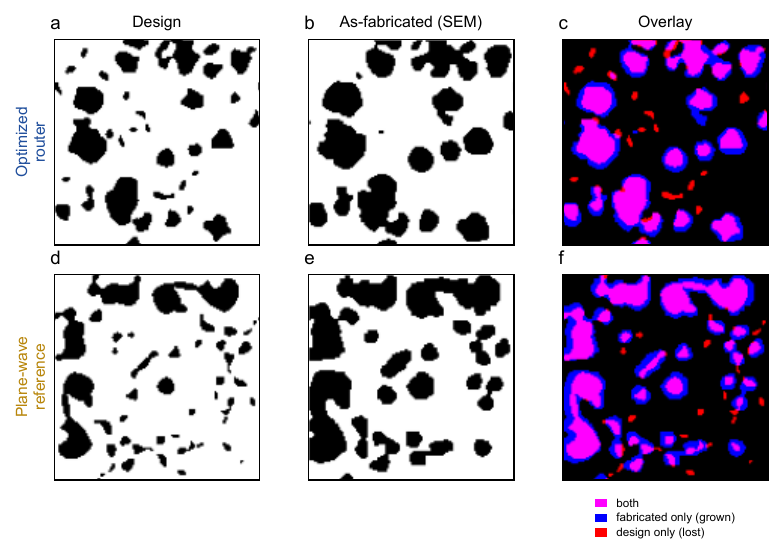}
\caption{\textbf{Design against the fabricated pattern.} For the optimized router
(\textbf{a--c}) and the plane-wave reference (\textbf{d--f}): the design layout,
the as-fabricated unit cell binarized from the SEM, and their overlap. In the
overlay, magenta marks structure present in both, blue the fabricated growth of
retained features, and red the design features that are lost. The lost structure
is concentrated in the sub-$90$\,nm features, which the fabrication does not
resolve.}
\label{fig:exp_asfab}
\end{figure}

The electron micrographs show that the smallest features of the design, below
about $90$\,nm, are absent from the fabricated film (Fig.~\ref{fig:exp_asfab}).
The retained features print with a small uniform growth, the regime the etch
variants of Sec.~\ref{si:integrity} bracket. The simulated spectra of
main-text Fig.~5(e,f) and of Fig.~\ref{fig:exp_intensity} use the as-designed
mask.

\end{document}